\documentclass{emulateapj}
\usepackage{graphics,graphicx,epsfig}
\usepackage{amsmath,amssymb,amsfonts} 
\usepackage{subfigure, multirow}
\usepackage{epstopdf}
\usepackage{natbib}
\newcommand{\lx}{$L_{\rm X}$}

\newcommand{\ergs}{erg s$^{-1}$~}
\newcommand{\ergss}{erg s$^{-1}$}

\newcommand{\lsun}{$L_{\odot}$}
\newcommand{\msun}{$M_{\odot}$~}
\newcommand{\Zsun}{$Z_{\odot}$}
\newcommand{\msunyr}{$M_{\odot}$ yr$^{-1}$~}

\newcommand{\eg}{\textit{e.g.,}~}
\newcommand{\ie}{\textit{i.e.,}~}

\newcommand{\HST}{\textit{HST}}

\newcommand{\Chandra}{\textit{Chandra}}

\newcommand{\hard}{2--10~keV}
\newcommand{\lhx}{$L_{\rm HX}$}

\defcitealias{L10}{L10}
\defcitealias{J11}{J11}
\defcitealias{M12}{M12}

\begin{document}

\title{Exploring the overabundance of ULXs in metal- and dust-poor local Lyman break analogs}
\author{
Antara R. Basu$-$Zych\altaffilmark{1,2},
Bret Lehmer\altaffilmark{1,3,4},
Tassos Fragos\altaffilmark{5}, 
Ann Hornschemeier\altaffilmark{1}, 
Mihoko Yukita\altaffilmark{1,9}
Andreas Zezas\altaffilmark{6,7,8}, 
Andy Ptak\altaffilmark{1}
}
\altaffiltext{1}{NASA Goddard Space Flight Center, Code 662, Greenbelt, MD 20771}
\altaffiltext{2}{Department of Physics, University of Maryland Baltimore County, Baltimore, MD 21250, USA}
\altaffiltext{3}{Department of Physics and Astronomy, The Johns Hopkins University, 3400 North Charles Street, Baltimore, MD 21218}
\altaffiltext{4}{Department of Physics, University of Arkansas, 825 West Dickson Street, Fayetteville, AR 72701}
\altaffiltext{5}{Geneva Observatory, University of Geneva, Chemin des Maillettes 51, 1290 Sauverny, Switzerland}
\altaffiltext{6}{Harvard-Smithsonian Center for Astrophysics, 60 Garden Street, Cambridge, MA 02138, USA}
\altaffiltext{7}{University of Crete, Physics Department \& Institute of Theoretical \& Computational Physics, 71003 Heraklion, Crete, Greece}
\altaffiltext{8}{Foundation for Research and Technology-Hellas, 71110 Heraklion, Crete, Greece}
\altaffiltext{9}{The Johns Hopkins University, Homewood Campus, Baltimore, MD 21218, USA}

\begin{abstract}
We have studied high mass X-ray binary (HMXB) populations within two low-metallicity, starburst galaxies, Haro 11 and VV 114. These galaxies serve as analogs to high-redshift ($z > 2$) Lyman break galaxies, and within the larger sample of Lyman break analogs (LBAs) are sufficiently nearby ($<$87 Mpc) to be spatially-resolved by \Chandra. Previous studies of the X-ray emission in LBAs have found that the \hard~luminosity per star formation rate (SFR) in these galaxies is elevated, potentially because of their low metallicities (12+$\log$[O/H]$=8.3$--8.4). Theoretically, the progenitors of XRBs forming in lower metallicity environments lose less mass from stellar winds over their lifetimes, producing more massive compact objects (\ie neutron stars and black holes), and thus resulting in more numerous and luminous HMXBs per SFR. In this paper, we have performed an in-depth study of the {\it only} two LBAs that have spatially-resolved \hard~emission with \Chandra~to present the bright end of the X-ray luminosity distribution of HMXBs (\lx$\gtrsim10^{39}$ erg s$^{-1}$; ultraluminous X-ray sources, ULXs) in these low-metallicity galaxies, based on 8 detected ULXs. Comparing with the star-forming galaxy X-ray luminosity function (XLF) presented by \citet{M12}, Haro~11 and VV~114 host $\approx$4 times more \lx$>10^{40}$\ergs sources than expected given their SFRs. We simulate the effects of source blending from crowded lower luminosity HMXBs using the star-forming galaxy XLF and then vary the XLF normalizations and bright-end slopes until we reproduce the observed point source luminosity distributions. We find that these LBAs have a shallower bright end slope ($\gamma_2=1.90$) than the standard XLF ($\gamma_2=$2.73). If we conservatively assume that the brightest X-ray source from each galaxy is powered by an accreting supermassive black hole rather than a HMXB and eliminate these sources from consideration, the luminosity distribution becomes poorly constrained but does appear to be consistent with a standard XLF. 
\end{abstract}


\section{Introduction}\label{sec:intro}

The X-ray view of high redshift ($z>2$) galaxies, such as Lyman break galaxies (LBGs), provides unique insight into the cosmic history of X-ray binaries (XRBs), compact object evolution, and black hole growth in galaxies. X-ray emission from LBGs is expected to significantly contribute to the heating of the intergalactic medium particularly in the early Universe \citep[IGM;][]{Mirabel11, Mesinger13, Pacucci14, Pober2015}. 

The 2--10~keV emission from star-forming galaxies is dominated by luminous accreting binary systems composed of massive (O- or B-type) stars accreting onto black holes or neutron stars, known as high mass X-ray binaries (HMXBs). This hard X-ray band is the one that we observe easily and directly in $z>2$ LBGs.

Our current knowledge of \hard~(2--10~keV) X-ray emission from distant LBGs is confined to the {\it average} detection of large numbers of objects, via the statistical ``stacking" technique \citep[][]{BrandtLyBreak,Nandra02, Seibert2002, Lehmer05, Laird06, Lehmer08, Cowie12, me-lbgstack}, with total stacked exposure times of months (to years) using the deepest X-ray surveys (\eg the \Chandra~Deep Fields; see \citealt{Xue11}). The few LBGs detected individually in the hard X-ray band at $z\approx3$ are typically dominated by active galactic nuclei (AGN) rather than star formation \citep{BrandtLyBreak,Nandra02,Lehmer05, Laird06, me-lbgstack}.~Extremely X-ray faint ``normal'' (\ie not AGN) LBGs are beyond the individual source detection limit of \Chandra; studies with {\it Athena} and other future large collecting area X-ray observatories may be able to reach these great distances ($z>1$). 

   \begin{figure*}[t!]
\begin{center}
  \includegraphics[height=2.5in]{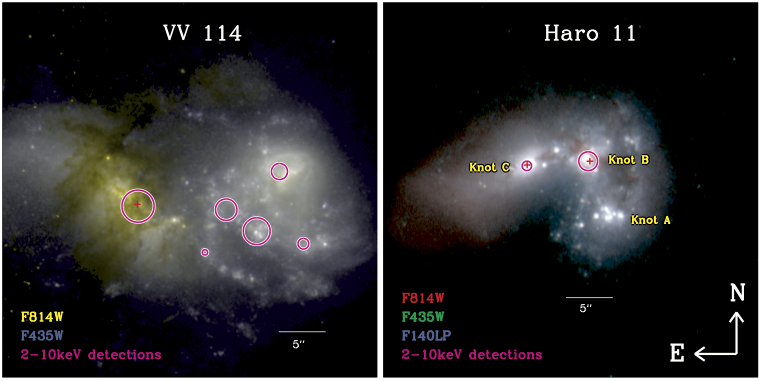}
    \end{center}
 \vspace{-0.15in}
  \caption{HST~image of our two LBAs: VV~114 (left), Haro~11 (right).
 ~Magenta circles mark the 2--10~keV source detections, with the circle size scaling with the X-ray luminosity. 
 The red crosses on VV~114 and Haro~11 mark sources that we consider as potential accreting SMBHs (see \S \ref{sec:AGN}).}
   \label{fig:images}
   \end{figure*}
The mode of star formation at earlier times in our Universe was very different on average than star formation observed in the local Universe.  Local galaxies, especially those with high star formation rates (SFRs), are generally relatively dusty \citep{Wang96, Hopkins2001, Afonso2003, Buat2005, Schmitt2006, daCunha2010,Garn2010} whereas high-$z$ populations like LBGs are dust-poor in comparison \citep{giarev,erb06, IglesiasParamo07, Reddy2008, Bouwens09}. 

In addition to the differences between modes of star formation in the past compared to the present (\ie relatively dust-poor LBGs versus dusty local star-forming galaxies), metallicity has significantly evolved over the history of the Universe. In the chemically pristine early Universe, generations of black holes formed in metal-poor environments. X-ray studies of nearby, dwarf metal-poor galaxies \citep{Mapelli09, kaaret, Prestwich2013, Brorby2014, Douna15} and of stacked $z>1$ LBGs \citep[][]{me-lbgstack, kaaret2014} offer hints that indeed the X-ray emission may be boosted in lower metallicity galaxies. This overall higher amount of the X-ray emission may be attributed to larger numbers of massive compact objects (neutron stars and black holes) and/or due to larger mass black holes (see papers on larger mass black holes in low-metallicity galaxies by \citealt{Prestwich07, Crowther2010} but see also \citet{Laycock2015a, Laycock2015b, Binder2015}).
   
Lyman break analogs (LBAs) are $z<0.3$ galaxies that resemble $z>2$ LBGs \citep{Heckman05, choopes, rod, me-ifu}. LBAs, selected by far-UV surface brightness ($>$10$^{9}$ $L_{\odot}$ kpc$^{-2}$), constitute a rare population of compact (half-light diameters $\approx$1--2 kpc), often disturbed systems, whose optical spectra indicate that they are starbursts. LBAs have low metallicities per stellar mass \citep{choopes} compared to the mass-metallicity relation observed in local galaxies \citep{Tremonti}. According to theoretical predictions, lower metallicities imply higher X-ray luminosities since more numerous and luminous HMXBs are produced per SFR \citep{Linden2010, F12}. Based on observations of 6 LBAs, \citet{me-chandralba} find that the hard (2--10~keV) X-ray emission (\lhx) per SFR is elevated in the LBA population, as expected given their low metallicities, and similar to the stacked LBG populations at $z=2$--3 \citep{me-lbgstack}. Comparing LBAs with other galaxy populations, this study finds that \lhx/SFR is anti-correlated with metallicity, following the predictions from X-ray population synthesis models \citep{Fragos13}. In addition, since these are relatively local galaxies, this study rules out that AGN activity contributes significantly to the elevated \lhx/SFR based on extremely constraining limits from high signal-to-noise optical spectra and X-ray spectral properties (see \S \ref{sec:AGN} of \citealt{me-chandralba}). Due to their unique properties (\ie low metallicities and high SFRs), LBAs constitute an ideal sample for studying the effects of metallicity on X-ray binary populations. 

VV~114 and Haro~11 are the only LBAs that have spatially-resolved \hard~emission with \Chandra, and therefore allow for more detailed study of their X-ray binary populations. Spatially resolving $z>2$ LBGs into individual luminous binaries would likely require \HST-quality spatial resolution (better than 0.1\arcsec, corresponding to physical sizes of $\sim$1~kpc) and a large leap in sensitivity (\ie collecting area) over what is currently available with the \Chandra~X-ray Observatory. 

In this paper, we advance our previous work on the {\it global} X-ray emission from LBAs and examine the spatial distributions and X-ray luminosity functions (XLFs) of luminous HMXBs in VV~114 and Haro~11 (\S\ref{sec:data}). In \S\ref{sec:results}, we present evidence that these galaxies appear to have more ULXs per SFR than expected for star-forming galaxies. Modulo source-blending effects (\S\ref{sec:confusion}), we test different XLF normalizations and bright-end slopes to fit the observations and find that 
the best-fit XLF has a shallower slope at the bright end than the standard XLF based on local star-forming galaxies \citep{M12}.  
While these galaxies have been carefully selected to avoid AGN based on their global properties (\eg optical and infrared emission lines), we discuss the potential contamination from AGN in our XLF in \S\ref{sec:AGN}. We present the physical interpretation for our results based on theoretical predictions in \S\ref{sec:phys}. In \S\ref{sec:LBAs}, we investigate how slope and normalization of the XLF can be parameterized by metallicity to describe the population of LBAs, and compare their effects of \lhx/SFR with other studies of low metallicity galaxies (\eg z$\sim$2 LBGs and local dwarf galaxies). We summarize our conclusions and include discussion about future research efforts in \S\ref{sec:conclusion}. 

Throughout this paper, we assume that the initial mass function (IMF) follows \citet{Kroupa} and have corrected SFRs to this assumption when comparing with other studies. Metallicities, henceforth, refer to gas-phase metallicities estimated using the oxygen (OIII $\lambda$5007) and nitrogen (NII $\lambda$6584) emission line ratios and the method described by Equation 3 in \citet[``PP04 {\it O3N2}'']{PP04}. As shown by \citet{KewleyEllison08}, metallicity values can vary systematically by $\sim0.7$~dex when comparing different measurement methods, and the PP04 {\it O3N2} is one of the most robust. 

\section{Sample Description and Analysis}\label{sec:data}
\setlength{\tabcolsep}{0.3em}
\begin{table}[t!]
\refstepcounter{table}\label{tab:source}
\begin{center}
{\bf Table 1:} Summary of Observations for Spatially-Resolved LBAs 
\scriptsize
\begin{tabular}{clccccc}
\hline
\hline
 & $ \alpha_{\rm J2000}$ & $\delta_{\rm J2000}$ & $D_{\rm L}$ & & $t_{\rm exp}$&$L_{\rm HX, limit}^{a}$\\
Name  & (hr) & (deg) &  (Mpc) & ObsID & (ks) & ($10^{39}$ erg s$^{-1}$) \\

\hline
VV~114 & 16.946 & -17.507 &  88 & 7063 & 59& 2.0 \\
Haro~11 & 9.219 & -33.555 &  86 &8175 &  54& 2.2\\
\hline
LBA~082355 & 125.979& 28.106 & 210 &  13012  &  9& 76 \\
\hline
\end{tabular}
\end{center}
$^{a}$ Estimate of the observing limit based on detecting 6 counts in the 2--10~keV band.

\end{table}

In this paper, the focus is to study the luminosity distributions of XRBs in the spatially resolved, and therefore nearest, LBAs. The subsample of two galaxies is drawn from the sample of six galaxies studied in \cite{me-chandralba}. We refer the reader to this paper for details on the sample selection. Briefly, LBAs are $z<0.3$ galaxies that satisfy the following FUV luminosity and surface brightness selection criteria: $L_{\rm FUV}>2\times10^{10}$~\lsun (FUV refers to 1500~\AA) and $I_{\rm FUV}>$10$^{9}$~\lsun kpc$^{-2}$. The only two LBAs that have spatially-resolved \hard~emission with \Chandra~are Haro~11 \citep[PI: J. Grimes]{Grimes07} and VV~114 \citep[PI: T. Heckman]{Grimes06}, permitting the study of their HMXB populations. 

Along with coordinates and redshifts, we provide a summary of the X-ray observations for these galaxies in Table \ref{tab:source}, including the \Chandra~observation ID (ObsID), exposure times ($t_{exp}$), and limiting 2--10~keV luminosity ($L_{\rm HX, limit}$). The limiting point source luminosity is based on having $>6$ counts in the 2--10~keV observation, which is the minimum number of counts observed for our detected sources, using the method described below. We provide this point source estimate for comparison purposes only and note it doesn't include detailed calculations for e.g., source confusion. 
The \Chandra~data were processed as described in \cite{me-chandralba}. Since we are interested in the XRB properties, we minimize contributions from the diffuse X-ray gas emission by confining our analysis to the 2--10~keV (hard-band) emission. Using \texttt{wavdetect} in CIAO version 4.4 with the significance threshold set to 10$^{-6}$ and wavelet scale sizes of 1.0, 1.4, and 1.8, we measure point source properties. We found that larger wavelet scales returned blended sources. The source intensities measured by \texttt{wavdetect} are generally not very accurate, especially in the case of blended sources embedded in diffuse emission. This is because \texttt{wavdetect} uses the wavelet scale that maximizes the significance of source, which may well include surrounding diffuse emission, or emission from neighboring sources.   In order to assess the accuracy of the source intensities measured by \texttt{wavdetect} we performed manually the source photometry, following the same approach as in \citet{Zezas06}.  For each source we defined 1\arcsec~radius circular apertures centered on the \texttt{wavdetect} positions. This radius includes at least 80\% of the encircled energy at 2--4~keV.  We also defined 1.5\arcsec~radius apertures used to excise the sources from the X-ray images, in order to create a source-free image used for the determination of the local background for each source. From this image we measure the background surface brightness around each source defining an annular region with an outer radius of at least 4\arcsec~(the inner radius is set by the excision region), taking care to sample in a representative way the local background of a source, while avoiding any other regions of diffuse emission. The source fluxes determined by this manual method are within errors of the \texttt{wavdetect} values. Ultimately, we chose to use the \texttt{wavdetect}-derived fluxes to maintain consistency with the method used for simulations presented later (\S\ref{sec:confusion}). We convert the measured \hard~count rate into fluxes in the same band by using \texttt{PIMMS} with a simple power law model with $\Gamma=1.9$ and Galactic column densities of $N_{\rm H}=3\times10^{20}$ cm$^{-2}$. Applying the distances given in Table \ref{tab:source}, we calculate luminosities from these \hard~flux measurements. Detected X-ray sources are shown in Figure \ref{fig:images} as magenta circles, whose size scales with \hard~luminosities. The numbers of detections having observed \lhx $>10^{40}$ \ergs  in each galaxy are given in Table \ref{tab:obs}.

\begin{table}[t!]
\refstepcounter{table}\label{tab:obs}
\begin{center}
{\bf Table 2:}  Properties of Spatially-Resolved LBAs
\scriptsize
\begin{tabular}{ccccccc}
\hline
\hline
&$\log M_\star$&SFR &  &  & \\
Name  & (\msun) &(\msunyr)&12$+\log$(O/H) & $N_{\rm exp}^{a}$  & $N_{\rm obs}^{b}$ \\

\hline
VV~114  & 10.6 & 38 & 8.4 & 1.6 & 5\\
Haro~11 & 9.84 & 11 & 8.3 & 0.4 & 2 \\
\hline
LBA~082355 & 9.5 & 17 & 8.2 & 0.6 & $>3^{\dagger}$\\
\hline
\end{tabular}
\end{center}
$^a$Number of sources with \lx$>10^{40}$\ergs expected based on star-forming galaxy XLF \citep{M12}. \\
$^b$Number of sources with \lx$>10^{40}$\ergs detected in the 2--10~keV band. \\
$^{\dagger}$ Because of the shallow observation of this target, we are unable to perform any analysis of the 2--10~keV emission in LBA~082355 and our lower limit estimate of the number of the HMXBs in this galaxy is based on the 0.3--10~keV emission. We mention this galaxy briefly in \S \ref{sec:LBAs}, but exclude it from our analysis. 
\end{table}

\section{Results}\label{sec:results}
\begin{figure}[t!]
  \begin{center}
    \includegraphics[height=2.9in]{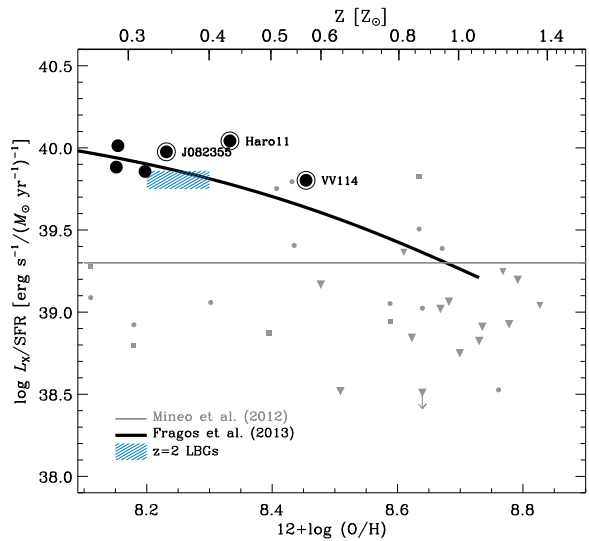}
              \caption{\lhx/SFR depends on gas phase metallicity, as predicted by XRB population synthesis models \citep[thick black line;][]{Fragos13}. This figure is modified from \citet{me-chandralba} to show the galaxies discussed in this paper, which appear as labelled encircled points. }\label{fig:metal}
            \end{center}
\end{figure}

\cite{me-chandralba} argue that lower metallicities in the LBA population may be driving the observed higher values of \lx/SFR. Figure \ref{fig:metal}, adapted from \citet{me-chandralba} to mark the specific galaxies discussed in this paper, shows how \lx/SFR decreases with increasing metallicity following the theoretical prediction \citep[black solid line;][]{Fragos13}. Furthermore, metallicity evolution appears to explain the mild evolution in \lx/SFR with redshift that is observed in LBGs in deep X-ray surveys between $z=1$--4 \citep{me-lbgstack}. Metallicity evolution is expected theoretically based on the application of XRB population synthesis models on mock galaxies from the Millenium simulation \citep{F12}. Theoretically, weaker stellar winds in low-metallicity environments allow more numerous and more massive stellar mass black holes to form, as well as more Roche-lobe overflow versus wind-fed systems, thereby producing more numerous and luminous HMXBs per SFR in metal-poor galaxies. Here, we use \Chandra~observations of the extended, low-metallicity LBAs to directly study the bright end of the HMXB populations in these galaxies. We focus on comparing how the HMXB populations in low redshift LBAs differ from more representative local, star-forming galaxies \citep[\eg][]{M12} to address why the population of LBAs and high-$z$ LBGs exhibit elevated \lx/SFR. 

\subsection{Excess of luminous XRBs}\label{sec:ULX}
Both HMXBs and low mass X-ray binaries (LMXBs) can contribute to the hard-band (2--10~keV) X-ray point source populations in galaxies. HMXBs trace young stellar populations (those with lifetimes $\lesssim10^{6-8}$ yrs), with X-ray emission proportional to the recent SFR; in contrast, the donors in LMXBs are stars with ages $\gtrsim10^{8-10}$ yrs and the X-ray emission from these sources traces past star formation (\ie they trace the stellar mass, M$_\star$). 

\cite{me-chandralba} describe the method that was used to measure metallicities, SFRs and M$_\star$ in this sample and we give these values in Table \ref{tab:obs}. Briefly, we use UV and infrared (IR) luminosities to measure SFRs from the relation given in \citet{Bell05} and 2MASS K$_S$ magnitudes with SDSS $u\prime-z\prime$ colors to estimate $M_\star$ according to \citet{Bell03}. 

Within each galaxy, we detected 2--6 X-ray sources.  The galaxies presented in this paper have specific SFRs (\ie SFR per stellar mass, sSFR$\equiv$SFR/M$_\star$) $>10^{-10}$ yr$^{-1}$, which suggests that HMXBs dominate the X-ray emission \citep{Colbert04, Lehmer2010}. Therefore, we will henceforth assume that the individual X-ray sources detected within VV~114 and Haro~11 are HMXBs. In \S\ref{sec:AGN}, we discuss the possibility that the X-ray emission in some of the sources may originate from accreting supermassive black holes (SMBHs). 

Given the relatively shallow observations (see Table \ref{tab:source}), individual XRBs are detected only at very high luminosities ($>10^{40}$ \ergs) meaning that they are all well above the threshold to be considered ultraluminous  X-ray sources (ULXs). By definition, ULXs are off-nuclear sources. In these morphologically irregular systems, we consider that all of the individual sources are ULXs. However, we revisit this assumption in \S\ref{sec:AGN}. 

   \begin{figure}[t!]
\begin{center}
\includegraphics[width=3.3in]{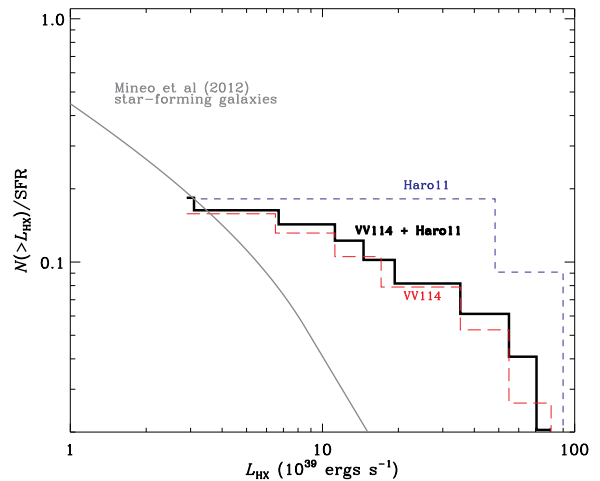}
\end{center}
\vspace{-0.15in}
\caption{We measure the cumulative number counts for VV~114 (red) and Haro~11 (blue) and compare with the star-forming galaxy XLF from \citet[shown in gray]{M12}. Based on our observations of these two LBAs (black curve shows the combined XLF from both galaxies), we find that low metallicity LBAs appear to have more ULXs at \lhx$>10^{40}$~\ergs than expected. }
\label{fig:xlf}
\end{figure}
\begin{table}[h!]
\refstepcounter{table}\label{tab:XRBs}
\begin{center}
{\bf Table 3:} HMXBs detected in LBAs
\scriptsize
\begin{tabular}{ccccl}
\hline
\hline
$\alpha_{\rm J2000}$& $\delta_{\rm J2000}$& &  $\log$\lhx & \\
(hr) & (deg) & Net Counts$^{*}$& (\ergs) & Notes \\

\hline
\multicolumn{5}{c}{VV~114 (ordered by decreasing \lhx)}\\
\hline

16.94795   &   $-$17.50711    &  179.0$\pm{15.1}$   & 40.79$\pm{0.04}$ & AGN?$^{a}$ \\
16.94472   &   $-$17.50776    &  142.3$\pm{13.5}$   & 40.68$\pm{0.04}$ & \\
16.94555   &   $-$17.50720    &  64.1$\pm{9.4}$   & 40.34$\pm{0.07}$ & \\
16.94410   &   $-$17.50621     & 36.0$\pm{7.2}$  & 40.09$\pm{0.10}$ & \\
16.94345   &   $-$17.50808     & 29.6$\pm{7.1}$  & 40.00$\pm{0.12}$ & \\
16.94613   &   $-$17.50830     &  8.1$\pm{3.0}$  & 39.46$\pm{0.25}$  & \\
\hline
\multicolumn{5}{c}{Haro11 (ordered by decreasing \lhx)}\\
\hline
9.218438   &   $-$33.55465   &   209.0$\pm{16.3}$  &  40.90$\pm{0.04}$ & Knot B; X1$^{a}$; AGN?$^{b}$ \\
9.219528   &   $-$33.55471   &   43.9$\pm{7.4}$  & 40.22$\pm{0.08}$  & Knot C; X2$^{a}$ \\
\hline
\end{tabular}
\end{center}
$^*$2--10~keV background-subtracted counts, based on {\texttt wavdetect} analysis. \\
$^a$Discussed in detail in \citet{PrestwichHaro11}\\
$^b$We treat this source as a potentially accreting SMBH in \S\ref{sec:AGN}
\end{table}

Our first assessment of the number of HMXBs detected in each galaxy is tabulated in Table \ref{tab:obs} (column 6), where we count the number of 2--10~keV sources with \lhx$>10^{40}$~\ergs found in each galaxy. In total, we detected 7 \lhx$>10^{40}$~\ergs point sources in VV~114 and Haro~11 (and an additional source with \lhx$=2.3\times10^{39}$~erg s$^{-1}$ in VV~114). We compare this value with the expected number of sources with \lhx$>10^{40}$~\ergs, based on the XLF derived from star-forming galaxies by \cite{M12} for the SFR of each galaxy (given in column 5 of our Table \ref{tab:obs}). Based on this comparison, we find that the total number of detected ULXs with \lhx $>10^{40}$~\ergs (7 sources) is $\sim 1.5$--10$\times$ higher than the expected number (2.0$^{+2.6}_{-1.3}$) for the entire sample. The Poisson probabilities for detecting at least the observed number of ULXs, based on the expected number for each galaxy (which is listed in column 5 of Table \ref{tab:obs}), are $P(N>2)=0.008$, $P(N>5)=0.006$ and $P(N>7)=0.001$ for Haro~11, VV~114 and both galaxies combined. We show this excess of high luminosity sources, normalized by SFR, in Figure \ref{fig:xlf} for Haro~11 (blue dashed line) and VV114 (red long dashed line), as well as for both galaxies combined (black solid line). Table \ref{tab:XRBs} lists the individually detected HMXBs (\ie their locations, background-subtracted counts, and 2--10~keV luminosities) for both galaxies. 

In the following section, we explore potential causes for this excess at the bright luminosity end. We start by considering whether the elevated number of luminous HMXBs is physical or artificially induced by source blending within unresolved star-forming regions within the galaxies. 

\subsection{Source Blending}\label{sec:confusion}

Are the detected sources truly ULXs with luminosities $> 10^{39}$\ergs or are these ``sources'' actually clusterings of several lower-luminosity HMXBs blended into single \Chandra~point sources? In this section, we address this question and assess the level at which source confusion, referring to the blending of several individual sources into a single unresolved higher luminosity ``source'', contributes to our analysis. To this end, we use the \texttt{MARX} version 5.0.0\footnote{Additional information is available at \url{http://space.mit.edu/ASC/MARX/}, including the publicly accessible suite of programs.} ray-tracing code to construct simulated images by placing fake sources in locations that are spatially populated following the light distributions of high spatial resolution \HST~images (see background images in Figure \ref{fig:images}), which act as proxies for star formation maps. To assign luminosities to these simulated sources, we draw probabilistically from the mean XLF presented by \cite{M12} for normal star-forming galaxies. 

The global SFR for each galaxy determines the total number of sources that will be simulated for that galaxy, as given by equation 20 in \cite{M12}. Our first step in the simulation is producing a SFR map. Ideally, we would use the UV $+$IR images to create SFR maps (see \citealt{Mineo14} for an example). Since the available UV and IR images have insufficient spatial resolution (\ie their PSFs are worse than \Chandra's) for this exercise, we use instead archived \HST~F814W optical images, acquired from the Hubble Legacy Archive, for SFR maps. While the F814W image does not trace star formation directly, the galaxies are expected to have similar morphology in this filter as they do in UV$+$IR images. Both galaxies have been observed with various \HST~filters, however we chose to use the F814W image for our analysis for consistency since this filter is common to both galaxies. We note that the other common filter is F435W, however the F814W filter is less affected by dust attenuation. As a check, we performed the simulation on all the available filters (F435W and F814W for VV~114 and F140LP, F435W, F606W, F814W for Haro~11) and found that the choice of filter makes little difference to the final simulated results and ultimately does not affect our results. 

Artificial sources are randomly placed according to the optical light distribution, drawing randomly from the luminosity distribution shown by the gray solid line in Figure \ref{fig:xlf}, which is given in equation 18 of \cite{M12}. Then we use \texttt{MARX}, which takes the \Chandra~PSF and instrument response into account using ray-tracing, to simulate the final \Chandra~image given the artificial sources. Using the \Chandra~observations for each galaxy, we estimate the background to add characteristic signal to our simulated image. 

   \begin{figure}[t!]
\begin{center}
\includegraphics[height=3.1in]{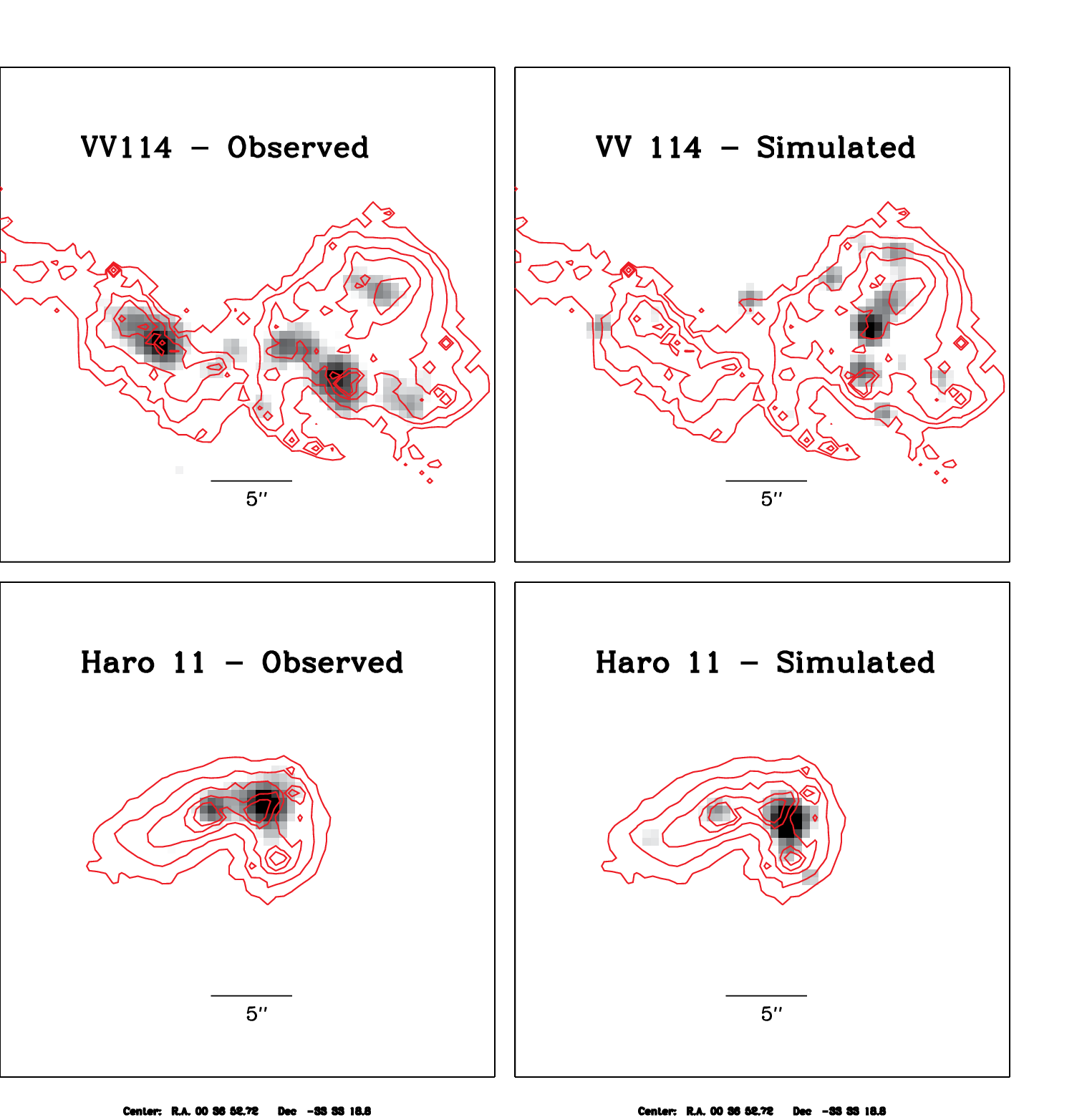}
\end{center}
\vspace{-0.15in}
\caption{We show the observed $\Chandra$ images (left) alongside with the MARX output images for the simulations (right), assuming X-ray luminosities distributed like typical star-forming galaxies from \citet{M12}, for VV~114 (top images) and Haro~11 (bottom images). Note that here we show only {\em one} realization, however we run 100 realizations of the simulations for our full analysis. In the simulations, X-ray sources are spatially distributed according to the SFR maps, which are estimated using the \HST~F814W filter and shown here as red contours. }
\label{fig:marx_image}
\end{figure}
At this point, the simulated image represents an observation of a star-forming galaxy with potentially blended sources with the same exposure as the actual observation. In Figure \ref{fig:marx_image}, we show one realization of the simulated image (right panels) compared to actual observations (left panels), for VV~114 (top) and Haro~11 (bottom). Finally, a simulated source list is constructed for each simulation using \texttt{wavdetect} with the same parameters as we used to analyze the actual data (see \S\ref{sec:data}). 

These steps are repeated 100 times, in order to minimize stochastic variations between separate realizations and compute errors on the simulated XLFs. The gray dashed line in Figure \ref{fig:xlf-sim} shows the effect that source blending has on altering the bright end of the XLF from the typical star-forming galaxy \citep[gray solid line;][]{M12}. 

The observed cumulative XLF, which combines the Haro~11 and VV~114 sources, is shown as a black histogram in Figure \ref{fig:xlf}. The errors on the cumulative XLF account for uncertainties on the source intensities as well as Poisson uncertainties on the number of sources in each bin following the approach of \citet{Zezas_antennae}. In short, for each source we estimated 1000 samples of its intensity assuming a Poisson distribution with a mean equal to its net number of counts. We defined coarse bins of unequal size centered at the intensity of each observed source. Each one of the 1000 sampled XLFs  was binned according to this binning scheme. The number of sources in each bin for each sample was determined by sampling from a Poisson distribution with a mean of 1 source. Then each one of these binned XLFs was converted to a cumulative XLF. From these 1000 cumulative binned XLFs we calculated the mean number of sources in each bin, and its standard deviation, which is our adopted uncertainty. 

\begin{figure}[t!]
\begin{center}
\hspace{-0.5in}
\includegraphics[width=3.7in]{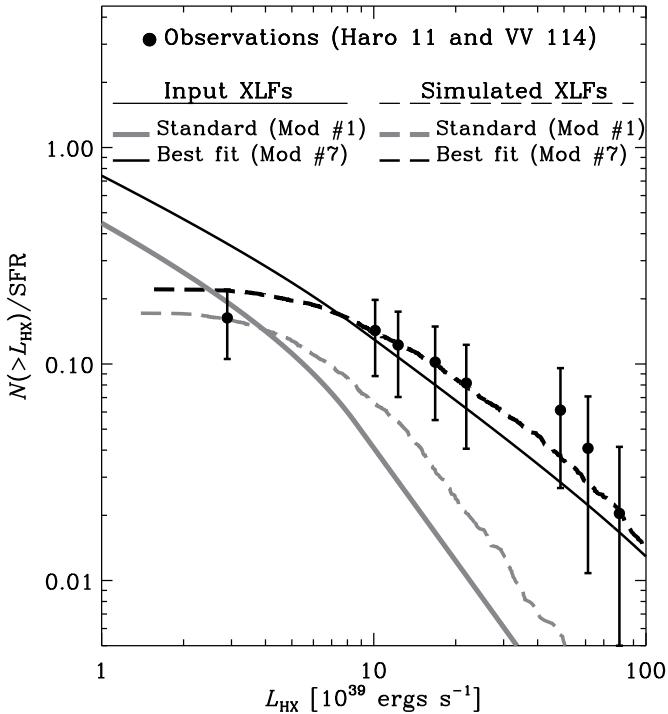}
\end{center}
\vspace{-0.15in}
\caption{We simulate source-blended XLFs (dashed lines) by drawing statistically from the input XLFs (solid lines) and constructing simulated images and XLFs. We show two cases here: gray curves for the standard model \citep[$\xi=1.95$ and $\gamma_2$=2.73;][]{M12} and black curves for our best-fit model (Model \# 7: $\xi=2.93$, 1.5$\times$ higher than the standard model, and $\gamma_2$=1.90). The combined observations for Haro~11 and VV~114 appear as solid points. Based on these simulations, the source-blended standard model is not a good representation of the data and is ruled out at $>99$\% confidence level by the KS test. }

\label{fig:xlf-simstd}
\end{figure}

Figure \ref{fig:xlf-simstd} shows that the observations (black points) are not consistent with a normal XLF that suffers from source blending (dashed gray line). Based on the Kolmogovrov-Schmirnoff (K-S) test, the standard XLF can be ruled out at $>99$\% confidence level (discussed in more detail in the following section). We find that the observed points still show an excess {\em above} the simulated data, possibly suggesting that a different XLF may better describe the HMXBs in these galaxies. We further explore this possibility in the following section. 

\subsection{The bright end of the XLF}\label{sec:fit}

   \begin{figure}[t!]
\begin{center}
\hspace{-0.5in}
\includegraphics[width=3.7in]{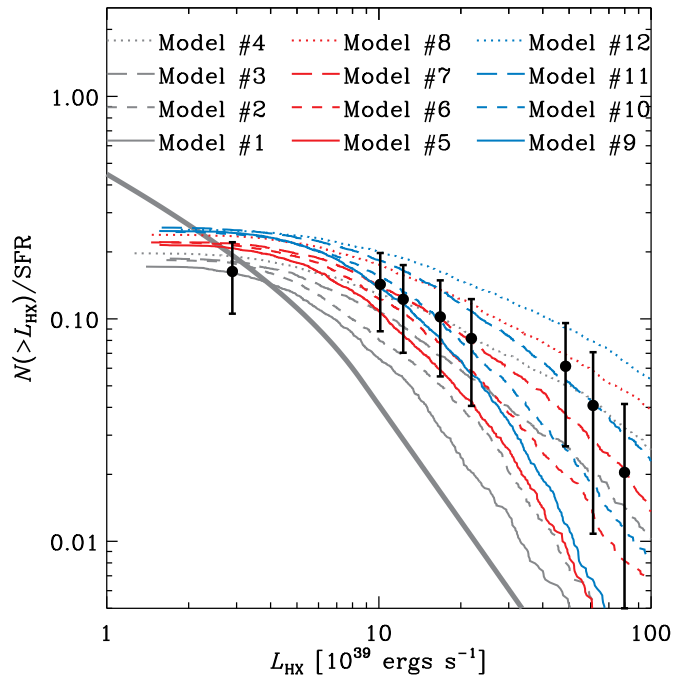}
\end{center}
\vspace{-0.15in}
\caption{As in Figure \ref{fig:xlf-simstd}, the black points show the observed XLF and the thicker gray line shows the input standard XLF (without source blending effects) in both panels. The simulated XLFs are displayed in different colors and line styles to show the effects of varying normalizations and bright-end slopes, respectively. Table \ref{tab:models} provides the parameter values that correspond to the models. Model \#1 refers to the standard star-forming galaxy XLF. }
\label{fig:xlf-sim}
\end{figure}

\begin{table}[t!]
\refstepcounter{table}\label{tab:models}
\begin{center}
{\bf Table 4:} Comparison of input XLF models 
\scriptsize
\begin{tabular}{ccccc}
\hline
\hline
& & & \multicolumn{2}{c}{C-Statistic} \\
\cline{4-5}
&  &  & All  & Excluding \\
Model \# & $\xi$ & $\gamma_2$ & Sources & AGN candidates \\

\hline
1  & 1.95 & 2.73 &    137  &    20.9  \\
2  & 1.95 & 2.30 &       137 &     20.6  \\
3  & 1.95 & 1.90 &      23.0  &    21.0\\
4  & 1.95 & 1.58 &      23.6  &    22.0 \\
5  & 2.93 & 2.73 &      26.6  &    18.8 \\
6  & 2.93 & 2.30 &      25.0  &    20.2 \\
7  & 2.93 & 1.90 &      22.8  &    19.6 \\
8  & 2.93 & 1.58&      24.6  &    23.4 \\
9  & $Z$-dep$^{a}$& 2.73 &      26.7  &    20.6  \\
10  & $Z$-dep$^{a}$ & 2.30 &      23.8  &    20.9 \\
11  & $Z$-dep$^{a}$& 1.90 &      23.0  &    22.3 \\
12  & $Z$-dep$^{a}$& 1.58 &      25.3  &    24.7 \\

\hline
\end{tabular}
\end{center}
$^{a}$ Metallicity-dependent normalization is set by the prediction by \citet{Fragos13}, which is $\xi$=4.83 for Haro~11 and 3.37 for VV~114. 
\end{table}
Rather than adopting the star-forming galaxy XLF given by \cite{M12} (henceforth referred to as the ``standard XLF''), in this section we ask:  can a different distribution of X-ray luminosities describe the observed detections more consistently for the LBAs? We follow the same prescription as described above, but use a grid of models that vary the slope of the bright end ($\gamma_2$) and the normalization ($\xi$) of the input XLF. 

The XLFs follow the form: 
\begin{equation}
dN/dL_X=\xi~SFR \times \begin{cases}
    L_{38}^{-\gamma_1},& L_{38} < L_b, \\
    L_b^{\gamma_2-\gamma_1} L_{38}^{-\gamma_2},              & L_b\geq L_{38} \geq L_{cut}, 
\end{cases}
\end{equation}
where $L_{38}$=\lx/$10^{38}$~erg s$^{-1}$, $L_b$ is the break luminosity, and $\xi$ is the average normalization. We use the following values from \citet{M12}: $L_b=110^{+110}_{-34}$, $\gamma_1=1.58\pm{0.02}$, $L_{cut}$=$5\times 10^3$, but we vary $\xi$ and $\gamma_2$. Our grid samples bright-end XLF slopes between the standard XLF \citep[$\gamma_2$=2.73; ][]{M12} and a ``no break'' scenario \citep[$\gamma_2=\gamma_1$=1.58;][]{M12} in four nearly-equal steps: $\gamma_2={2.73, 2.30, 1.90, 1.58}$. The ``no break'' case explores a scenario similar to the one observed for the Antennae pair of galaxies \citep{Zezas_antennae}. \citet{Zezas_antennae} conclude that the lack of break implies that the ULXs in the Antennae galaxies are not entirely a new population (\eg intermediate mass black holes, IMBHs) but, mainly, a luminous extension of massive ($M_{BH}\sim 80M_{\sun}$) HMXBs accreting at the Eddington limit. 

We sample only a few normalizations\footnote{In practice, the values of the normalizations are actually $\xi$/1.6. In order to compare the parameters with \cite{M12}, we state the ``normalization'' as $\xi$. Since the units for the normalization are given in (\msunyr)$^{-1}$, the factor of 1.6 comes from converting the SFR-normalized XLF from a Salpeter IMF to Kroupa.}: one that is equivalent to the standard XLF \cite[$\xi=1.95$;][]{M12}\footnote{The quoted value for $\xi$ in \citet{M12} is 1.49. However, $\xi=1.95$ is the value used to derive their Equation 20 and is the more appropriate value to use here. (S. Mineo, private communication).}, one that is 1.5$\times$ higher than the standard XLF ($\xi=2.93$) and a final one that is metallicity dependent and based on the theoretical metallicity enhancement of \lhx/SFR versus metallicity ($\xi=4.83$ for Haro~11 and 3.37 for VV~114). The metallicity enhancement is measured using the prediction for X-ray luminosity per SFR made by \cite{Fragos13}, shown as a gray solid line in Figure \ref{fig:metal}, based on XRB population synthesis models, compared to that observed by \cite{M12} ($\log L_{\rm X}/$SFR$\sim39.3$ \ergs$(M_\sun$ yr$^{-1})^{-1}$; solid black line) given the galaxy's metallicity (provided in Table \ref{tab:obs}). We note that the \cite{Fragos13} models give a prediction about how the integrated emission varies with metallicity but does not give any information about the underlying distribution and whether the normalization or the shape of the XLF is changing. Here, we are making the assumption that the increase in \lhx/SFR is due to a higher normalization value. While the first value of $\xi$ tests variations of the slope assuming a local star-forming galaxy, the final normalization value tests a scenario where low metallicity galaxies have intrinsically more HMXBs at all luminosities. The middle value for $\xi$, which is not physically motivated, allows us to test the effect of an intermediate normalization on final observations for comparison.

Due to computational limitations, a finer grid of XLF model parameters was unfeasible. Using the grid of 12 unique models for each galaxy and the method described above, we simulate the source-blended XLFs.  We applied the Cash \citep[\texttt{cstat};][]{cstat} statistic within the \texttt{sherpa} package in \texttt{ciao} to determine the best model for both of the galaxies. Specifically, we compared the observed numbers of detected sources within some luminosity bin in log units($dN/d\log L_{\rm X}$) with the output $dN/d\log L_{\rm X}$ from our \texttt{MARX} simulation of blended sources given the input models described above to test the effects of source blending for different XLF inputs. The bin size that we used, $d(\log L_{\rm X})=0.07$ dex, corresponds to $\sim 1$ count ($\approx 3 \times 10^{38}$\ergs) around the detection limit of our observations, given in the last column of Table \ref{tab:source}. The advantage of using the Cash statistical measure is especially evident for cases such as ours where the smooth distribution of simulated model values is compared to data in which most of the luminosity bins have no sources while a few have a single detected source. 

We display the simulated data of potentially blended sources in the Figure \ref{fig:xlf-sim}, showing the effects of varying normalizations ($\xi=$1.95, 2.93, and $Z$-dependent) with the different colors (gray, red, blue, respectively) and bright-end slopes ($\gamma_2=$2.73, 2.30, 1.90, 1.58) with different line styles (solid lines, dashed, long-dashed, and dotted, respectively).  From the Cash statistic (C), the best model for the Haro~11 and VV~114 datasets is Model \#7: the model with $\gamma_2=1.90$ and the 1.5$\times$ higher normalization than the star-forming galaxies ($\xi=2.93$). We display this model in black in Figure \ref{fig:xlf-simstd}, solid lines showing the input XLF and dashed lines showing the effect of source blending on this model. 
By performing a two-sample KS test we find that the observed XLF is inconsistent with a parent distribution described by models \# 01 (standard XLF) and 02 at above the 99\% confidence level.  Since the KS test does not take into account the total number of sources, which however is a critical parameter in the description of the XLF, we also calculate the probability that from each one model we can obtain the observed number of sources down to a luminosity of $10^{40}$ \ergss, assuming the Poisson distribution. We limit ourselves to sources brighter than  $10^{40}$ \ergs since from Table 3 we see that the faintest source has very large uncertainty in its luminosity (it is not even a 3$\sigma$ detection). This analysis showed that we can reject models \#1 (the standard model), 2, and 11 at the $>90$\% confidence level on the basis of the expected number of counts, albeit with limited statistics at the bright-end where the difference between models is maximized. 
 It is noteworthy that \cite{M12} find that NGC3310, another low-metallicity galaxy \citep[$Z=0.4$\Zsun;][]{deGrijs03a, deGrijs03b}, deviates from the mean XLF by $\sim 2\sigma$, also exhibiting a flatter slope than the rest of the sample of star-forming galaxies. This galaxy also has an \lx(XRB)/SFR value that is a factor of $\approx$5 times higher than average \citep{Lehmer15}.

Before discussing the physical interpretation for these results, we address possible contribution from AGN on the observed XLFs.

\subsection{Caveats: Potential AGN}\label{sec:AGN}

Several studies have concluded that AGN do not contribute significantly to the global emission in Haro~11 \citep{Grimes07, Cormier2012} and VV~114 \citep{Grimes06}. In these galaxies, star formation related processes dominate the total emission in the infrared, optical and ultraviolet. Both of these galaxies reside firmly in the ``star-forming'' region of the BPT diagram \citep{BPT,VO87,Kauffmann03AGN}, which is a useful diagnostic for distinguishing ``AGN'' from ``star-forming galaxies'' based on emission line ratios ([OIII] $\lambda$5007/H$\beta$ vs. [NII] $\lambda$6584/H$\alpha$; shown explicitly in Figure 4 from \citealt{me-chandralba}). \citet{Jia} study the X-ray emission within a different population of $z=0.1$--0.3 LBAs, those that occupy the ``composite'' rather than the ``star-forming'' region of the BPT diagram, and find these to have \lhx$>10^{42}$ erg s$^{-1}$. They argue that these {\it composite} LBAs likely harbor Type 2 AGN but that star-formation dominates the bolometric luminosity in these galaxies, potentially mimicking the growth of SMBHs within dense, stellar clumps that is believed to occur within galaxies in the early Universe ($z>3$).

It is plausible that hidden AGN may be present, since most studies have not investigated individual localized regions within the LBAs to discern whether any are consistent with accreting SMBHs. For Haro~11, \citet{James2013} have detailed spectral analysis of the different knots and refer to them as super star clusters, and do not discuss the possibility that AGN lie within these regions. Recently, \citet{PrestwichHaro11} published detailed a X-ray study of the two X-ray sources found in Haro~11 and do consider the possibility that X-1 (coincident with Knot B and shown as a cross in Figure \ref{fig:images}) may be an AGN, but also argue that this source is consistent with being an intermediate mass black hole ($M_{BH}>$7600\msun) in the low-hard state. \citet{Iono13} used ALMA data to argue that a highly obscured AGN most plausibly produces the observed high ratio of HCN (4--3) to HCO$^+$ (4--3) molecular emission lines ($R_{\rm HCN/HCO^+}= 1.34\pm 0.09$), which also appear to be broad ($\sim$290 km s$^{-1}$), within the eastern nucleus of VV114 \citep[defined as the position of the peak emission in the Ks-band; see also][ also shown as a cross in Figure \ref{fig:images}]{Saito2015}.  

If we conservatively speculate that X-ray sources coincident with the optical nuclei of each galaxy are indeed obscured AGN (bearing in mind that these galaxies appear as disturbed systems, possibly the mergers of two galaxies, each containing a SMBH), we consider two possible obscured AGN in Haro~11 and one in VV~114 (Table \ref{tab:agntab} provides details about each source). We only consider one potential AGN in VV114 because peak nuclear optical emission in the Western component of VV114 does not coincide with any X-ray source. We mark these potential AGN candidates with crosses in Figure \ref{fig:images}. The numbers of 0.3--7.0~keV net counts for these sources range from 292--610 counts, therefore detailed spectral fitting is not possible. However, we perform a simple analysis: placing 1.5\arcsec~source apertures around the sources and subtracting background using 10\arcsec~apertures located 20\arcsec~away from the galaxy. Using \texttt{XSPEC} with the maximum-likelihood statistics set to \texttt{CSTAT}, we fit the resulting background-subtracted spectrum with an absorbed power law that {\it only} includes the Galactic column density (${N_{\rm H}/10^{20}~{\rm cm}^{-2}=1.3}$ and 2.0 for VV~114 and Haro~11, respectively) and find $\Gamma$ values ranging from 0.2--1.7 (see Table \ref{tab:agntab}). Additional spectral analyses of these sources is beyond the scope of this paper, but detailed analyses of these spectra can be found in \citet[Knots B and C in Haro11]{PrestwichHaro11} and \citet[VV114E]{Grimes06}. Since the observed spectra in this method includes emission from a diffuse background, the spectra of the point sources may be flatter than the fits imply. We note however, that flatter slopes do not necessarily prove that the sources are AGN. ULXs with significant obscuration, for example, those located within intense star-forming regions, could be consistent with these spectral constraints as well \citep[\eg the central ULX in NGC~253; ][]{Lehmer13}.

\begin{table}[t!]
\refstepcounter{table}\label{tab:agntab}
\begin{center}
{\bf Table 5:} AGN or HMXB? Spectral fitting constraints for potential AGN
\scriptsize
\begin{tabular}{cccccc}
\hline
\hline
X-ray Source &  RA & Dec &  Counts$^{a}$ & $\Gamma^{b}$ \\
\hline
VV~114 E  &16.94795   &   $-$17.50711 &  292 (290.5) & 0.2$\pm{0.1}$\\
Haro~11 (Knot B) & 9.218438   &   $-$33.55465 & 610 (608.5) & 1.1$\pm{0.1}$\\
Haro~11 (Knot C) & 9.219528   &   $-$33.55471 & 299 (297.7) & 1.7$\pm{0.1}$\\
\hline
\end{tabular}
\end{center}
$^{a}$ Net (Background-subtracted) counts for 0.3--7.0~keV data.\\
$^{b}$ Absorbed power law fit ({\texttt{XSPEC wabs $\times$ powerlaw}}) with Galactic absorption only (for VV~114: $N_{\rm H}=1.3\times10^{20}$cm${^{-2}}$; for Haro~11: $N_{\rm H}=2.0\times10^{20}$cm${^{-2}}$).  
\end{table}

   \begin{figure}[t!]
\begin{center}
\hspace{-0.5in}
\includegraphics[width=3.8in]{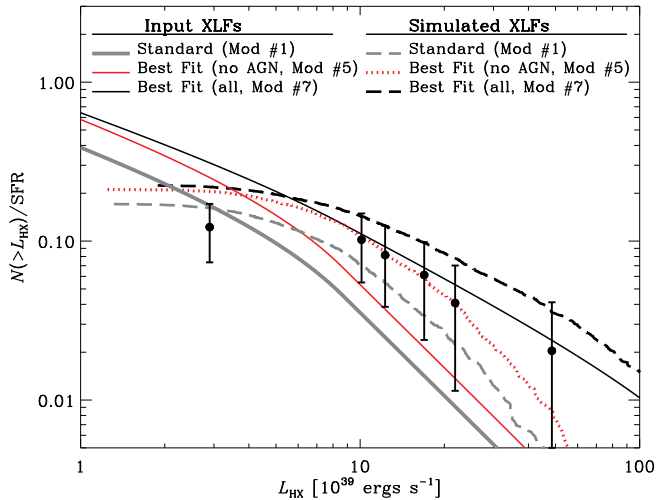}
\end{center}
\vspace{-0.15in}
\caption{Same as bottom panel of Figure \ref{fig:xlf-sim}, except that we have removed any potential AGN from both the observed points (filled circles) and simulated data (dashed curves). The black dashed line was the previous best fit (Model \#7, same as Figure \ref{fig:xlf-sim}), which is no longer the best description, but is a decent representation (second best model, see Table \ref{tab:models}). In addition we have added the red dotted line to show the best fit to these data: Model \#5 ($\xi=2.93$ and $\gamma_2=2.73$, same as the standard model), input XLF is shown as a solid red line. We note that our conservative analysis removed the sources at the bright end and therefore the models are not as well-constrained. }
\label{fig:AGNcorr}
\end{figure}
Based on the steep ($\Gamma=1.7$) slope for Haro~11 (Knot C), we decide this source is consistent with ULX spectra and do not consider this a potential obscured AGN \citep[see also ][]{PrestwichHaro11}. We note that {\it unobscured} AGN may have $\Gamma=1.7$, but optical and infrared emission line constraints rule out such AGN in Haro~11. Treating the remaining two X-ray sources (one from Haro~11 and one from VV~114) as AGN rather than XRBs, we proceed to correct our previous XLF, shown in Figure \ref{fig:AGNcorr}. 

We eliminate these sources to produce an ``AGN corrected'' observed XLF (black points). We also show new source blended simulations for a luminosity distribution given by the standard XLF (gray dashed curve) and that which was our previous best fit (Model \#7, black dashed curve; see \S\ref{sec:fit}: $\gamma_2$=1.90, $\xi=$2.93). 
These simulated curves follow the prescription described in \S\ref{sec:confusion}, but mask regions (using circular masks of 0.5\arcsec~radii) at the assumed AGN locations to produce new spatial distributions. In Haro~11, this masking blocks 9\% of the optical light, whereas in VV114, only 0.4\% of the optical emission is masked. In this case, we arrive at the best-fit simulated XLF (dotted red curve), which is based on the Model \#5 XLF (with $\xi=2.93$ and the same slope as the standard XLF; shown as a red dotted line).  Based on this analysis, we find that if the brightest sources in these galaxies are truly AGN, then the observed luminosity distribution and numbers of ULXs, can be explained by source blending HMXBs drawn from the standard XLF (KS test cannot rule this out at the 99\% confidence level; see Table \ref{tab:models}). 

We emphasize that we have taken the approach of performing a very conservative treatment of potential AGN, yet the results in this case are not very constraining.  Since the models diverge with increasing luminosity, losing two data points from the bright end of the luminosity distribution means that the constraints on these models are even weaker than in the previous fit (\S\ref{sec:fit}).  

\subsection{Physical Interpretation}\label{sec:phys}

We have studied how HMXB populations compare between star-forming galaxies and low-metallicity LBAs to explain the observed elevated \lhx/SFR in low metallicity, spatially-resolved galaxies. Many other studies have observed a surplus of ULXs in metal-poor galaxies \citep{Mapelli09, Mapelli2010, Mapelli2011,Kaaret2011,kaaret2014,Prestwich2013,Brorby2014,Douna15}. These studies, in addition to compounding evidence that LBAs and LBGs have elevated \lx~per SFRs \citep{me-lbgstack, me-chandralba}, suggest that metallicity influences the bright-end of the XRB luminosity function in an important way. In addition, the consistency between observations and XRB population synthesis models in these galaxies, spanning the past 12.2 Gyr of cosmic history \citep{me-lbgstack, me-chandralba}, further offers theoretical insight into the role that metallicity plays in the formation and evolution of HMXBs. In this study, we find evidence that the XLFs in our sample of low-metallicity galaxies have shallower slopes at the bright end ($\gamma_2=1.90$) compared to what has been observed for typical star-forming galaxies ($\gamma_2=2.73$)

\begin{figure}[t!]
\begin{center}
\hspace{-0.5in}
\includegraphics[width=3.8in]{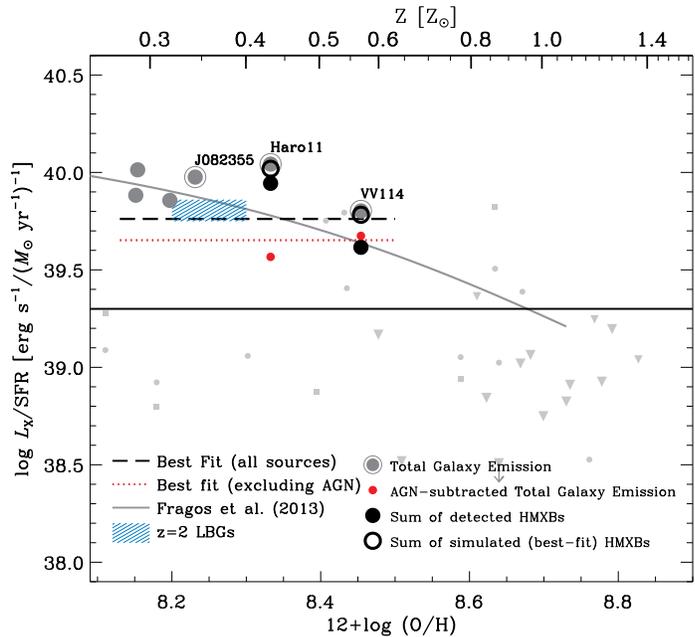}
\end{center}
\vspace{-0.15in}
\caption{Starting with Figure \ref{fig:metal}, we now add data based on current analysis of sources within Haro~11 and VV~114. The previous black points (now dark gray) show the global 2--10~keV emission for the galaxies, while the black points show the sum of luminosities from the detected HMXBs within each galaxy. The open black circles mark the sum of luminosities summing the luminosities by using the luminosity distribution of our best fit (Model\# 2), demonstrating that source-blended sources drawn from a flatter XLF distribution than the standard XLF also match the global emission and the \citet{Fragos13} prediction. The smaller red points show the effect of removing potential AGN sources from the observed detections. The lines provide the estimate of \lhx/SFR based on integrating the best-fit XLFs down to $10^{36}$\ergs for all the sources (black dashed, \S\ref{sec:fit}) and excluding AGN candidates (red dotted, \S\ref{sec:AGN}).}
\label{fig:metalsim}
\end{figure}

In Figure \ref{fig:metalsim}, we compare the different measurements of \lhx/SFR based on global emission from the galaxies (encircled gray points), sum of detected sources in the 2--10~keV \Chandra~observation (black solid point), sum of simulated source-blended sources assuming our best-fit XLF (open circle) and sum of detected sources without including our conservative assessment of potential AGN (addressed in \S\ref{sec:AGN}, red points). We note that the sum of the luminosities in the detected sources does not account for the total global emission, especially in VV~114, because undetected HMXBs produce additional 2--10~keV emission (which is also evident in the \Chandra~images shown in the left panels of Figure \ref{fig:marx_image}). The gray solid lines show the \citet{Fragos13} theoretical prediction, which is based on X-ray binary population synthesis models \citep[\texttt{StarTrack};][]{startrack} combined with cosmological simulations \citep[Millenium II;][]{MS-II}. Figure \ref{fig:metalsim} illustrates that the two galaxies studied in our paper, simulated with a flatter slope than the standard XLF, match the \citet{Fragos13} prediction fairly well. The black dashed line gives the \lhx/SFR expected based on integrating our best-fit XLF down to $10^{36}$\ergss, and shows good agreement with the other LBAs (gray points) and the stacked LBGs (blue shaded region). The red dotted line gives a similar result for the best-fit XLF measured from excluding potential AGN, predicting values of \lhx/SFR that are higher than those for star-forming galaxies (black solid line).

Studying the HMXB populations within galaxies with a range of metallicities, \citet{Douna15} also find evidence that \lx/SFR depends inversely with metallicity and the main cause for this is that low-metallicity galaxies (12+$\log$[O/H]$\lesssim$8.0, based on a different but roughly comparable metallicity estimate from ours) have a higher number of HMXBs/SFR. A weaker trend is that low-metallicity galaxies also have more {\it luminous} HMXBs. In terms of the luminosity function, the stronger effect should be the normalization of the XLF, with a secondary effect that is consistent with a flatter XLF slope. Similarly, \citet{Brorby2014} measure the XLF from 25 metal-poor (12+$\log$[O/H]$<$8.0) blue compact dwarfs (BCDs) and find a normalization $9.7\pm{3.2}$ higher than the standard XLF. While our results do not appear to show the factor of $\sim$10 enhancement in the normalization as these other works, the metallicities of our sample are somewhat higher (12+$\log$[O/H]=8.3--8.4) than their ``low-metallicity'' cases (12+$\log$[O/H]$<$8.0). Instead, our sample of slightly sub-solar metallicity galaxies appears to have a stronger effect on the bright-end slope. Having a similar metallicity to our sample \citep[$Z=0.4$\Zsun;][]{deGrijs03a, deGrijs03b}, NGC3310 appeared as a 2-$\sigma$ outlier from the typical XLF found by \citet{M12} in their sample, also exhibiting a flatter slope (at \lhx$\lesssim10^{40}$ erg s$^{-1}$) than the standard XLF. 

Although detailed modeling of metallicity effects on the XLFs for ULXs has not been done yet (see discussion in Linden), several theoretical arguments have been suggested in the literature to explain the increased \lhx/SFR with decreasing metallicity:
\begin{enumerate}
\item{Lower metallicity massive stars have weaker wind mass-loss during their evolution and reach the core-collapse phase more massive, on average. This results in a more numerous and more massive black hole population, and thus in a more luminous HMXB population \citep{Linden2010, Fragos12, Fragos13}.}
\item{Wind mass-loss leads to angular momentum loss from the orbit and thus to an orbit expansion. Hence, the lower metallicity binaries will have, on average, tighter orbits which will result in an increased number of Roche-lobe overflow (RLO-HMXBs) versus wind-fed systems. Since the former can drive much higher accretion rates, lower metallicity HMXB populations are expected to be more luminous \citep{Fragos12, Fragos13}.}
\item{All RLO-HMXBs go through a common envelope phase before the formation of the compact object. According to \citet{TS2000}, common envelope (CE) phases initiated while the donor star is in the asymptotic giant branch (AGB) will most likely lead to a merger, as AGB stars have not yet developed a clear core-envelope boundary. Hence, in order for a binary to survive the CE, the latter must be initiated while the radius of the donor star is larger than the maximum AGB radius and smaller than the maximum radius of the star in the super-giant phase. The maximum super-giant radius is not strongly dependent with metallicity, in contrast to the maximum AGB radius which decreases steeply with metallicity. Therefore, the range of initial orbital separations that can lead to survivable CE events increases strongly with decreasing metallicity, and so does the formation of luminous RLO-HMXBs \citep{Linden2010}.}
\item{An alternative explanation has been recently put forth by \citet{Justham15} who suggests that since low metallicity massive stars tend to expand later in their evolution compared to high metallicity ones, they will have developed a higher mass core by the time they go into the CE phase. These higher mass cores will in turn result in an increased number of black holes formed.}
\end{enumerate}

A second class of luminous HMXBs are wind-fed HMXBs with (super)giant donor stars. Given their current, very wide orbits, these systems are not expected to have undergone a CE phase prior to the compact object formation. This class of systems is not expected to have the same metallicity dependence as RLO-HMXBs. The metallicity dependence in these cases is complicated by two competing effects which contribute to higher X-ray luminosities. In higher metallicity regions, stronger winds from donors drive higher accretion rates. Lower metallicity systems can have closer orbits (i.e., the maximum radius during the Hertzsprung gap stage is smaller), which lead to more luminous HMXBs. Therefore the number of bright HMXBs peaks for intermediate metallicities \citep[$Z\sim0.2$\Zsun;][]{Linden2010}. However, we should note here that wind-fed HMXBs are not expected to reach super-Eddington luminosities \citep{Val2010, Wong2012, Wong2014}, in contrast to RLO-HMXBs with either black hole or neutron star accretors that can easily drive mass-transfer rates in excess of $\sim$10 times the Eddington limit \citep{Pod03, Rapp05, Fragos15a, Fragos15b}. Hence, we expect that ULXs at sub-solar metallicities are a population of bright HMXBs that experience mild super-Eddington accretion and undergo stable Roche-lobe overflow. This is the most consistent explanation for observations of ULXs thus far \citep[\eg][]{Gladstone09, Bachetti2013, Walton13, Walton14, Rana15}.

Specific to our study, \citet{Linden2010} predict that the number of ULXs increases by a factor of 2.5 between $Z=$ \Zsun~and $Z=$ 0.4\Zsun~(\ie the metallicity for Haro~11; see Figure \ref{fig:metal}). We note that this is close to the increase predicted by \citet[factor of 2.7]{Fragos13}, shown in Figure \ref{fig:metal}. \citet{Linden2010} find that the metallicity-dependence on the number of ULXs evolves with the age of the stellar population (time since burst of star formation). Specifically, the metallicity dependance is complex between 5--10~Myr post-burst, peaking at $Z=0.4$\Zsun. Then from $\approx$10~Myr after the star formation episode, the trend between ULX number and lower metallicities emerges.

In light of the theoretical discussion above, Haro~11 provides an interesting example. In Haro~11, high-spatial resolution multi-band photometric \citep[using \HST~imaging in 8 optical bands and VLT NaCo adaptive optics K-band imaging;][]{Adamo2010} and spectroscopic \citep[using the VLT/FLAMES optical integral field unit;][]{James2013} studies estimated the ages of three most prominent star forming regions (see Figure \ref{fig:images} for reference): Knot A is $4.9\pm0.4$~Myr, Knot B is $4.3\pm0.5$~Myr and Knot C is $7.8\pm0.3$~Myr. Therefore the ages for the star forming regions within Haro~11 fall within $\approx$5--10~Myr and, given that the galaxy's metallicity is $Z=0.4$~\Zsun, are consistent with the predictions presented above. The age range for Knot C is consistent with the RLO--HMXB pathway for HMXBs found in low metallicity environments. However, Knots A and B apparently have slightly younger ages, but also slightly $higher$ metallicities \citep[12$+\log$(O/H) = 8.0, 8.2 for Knots A and B, respectively, compared to 7.8 for Knot C;][]{James2013}. Modulo source blending, the simulated and \Chandra-detected sources do coincide with these star forming regions (Knots B and C). We note, however, that our source blending simulation using the metallicity-dependent XLF uses the average metallicity for this galaxy (given in Table 2) without accounting for the variation in metallicities of the different star-forming regions since the metallicities for individual star-forming regions is not well measured, in general. For example, no study has been published thus far estimating metallicities for different regions within VV~114.  

The relationship between metallicity and populations of ULXs is complex and depends on the age of the stellar population and is linked to the XRB formation pathway. In our one example where we have sufficient information to investigate this relationship, we find good agreement with the theoretical predictions by \citet{Linden2010}. LBAs, by definition based on their UV-selection, are some of the youngest galaxies in the local Universe and the LBAs in this paper contain stellar populations younger than $\lesssim 10$~Myr. The fact that these LBAs host large numbers of ULXs offers direct evidence for an excess of ULX rates at $<20$~Myr timescales. Based on our source blending simulation, we find that the observed bright end of the XLF in these LBAs arises from multiple blended HMXBs, which have a luminosity distribution that is flatter than the standard XLF. Therefore, the ultraluminous ``sources'' in these LBAs are better described as blended HMXBs, rather than an excess of individual ULXs. 

\subsection{Applying the full LBA sample}\label{sec:LBAs}
\begin{figure}[t!]
\begin{center}
  \includegraphics[height=2.3in]{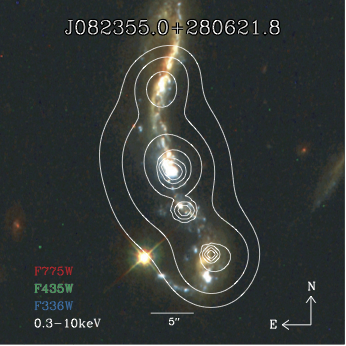}
    \end{center}
 \vspace{-0.15in}
  \caption{HST~image of LBA 082355.  
Due to the shallow observation (9ks; $L_{\rm HX, limit}=7.6 \times 10^{40}$ ergs s$^{-1}$) of LBA 082355, 0.5--8~keV X-ray contours are overlaid in white for this galaxy. }
   \label{fig:im_LBA3}
   \end{figure}

In addition to these two LBAs, LBA~082355 is also relatively nearby, at a distance of 210 Mpc, and shows spatially extended X-ray emission (see Figure \ref{fig:im_LBA3}. The current available \Chandra~data for this galaxy is insufficiently deep (9~ks exposure; 2--10~keV luminosity limit of 7.6$\times10^{40}$\ergss) to detect any sources in the 2--10~keV band or perform any detailed investigation into their XRBs. In Figure \ref{fig:metalsim} we show that the best-fit (Model \#7) integrated down to $10^{36}$\ergs describes \lhx/SFR better for LBA~082355 and the other LBAs than the standard XLF (Model \#1).

   \begin{figure}[t!]
\begin{center}
  \includegraphics[width=3.4in]{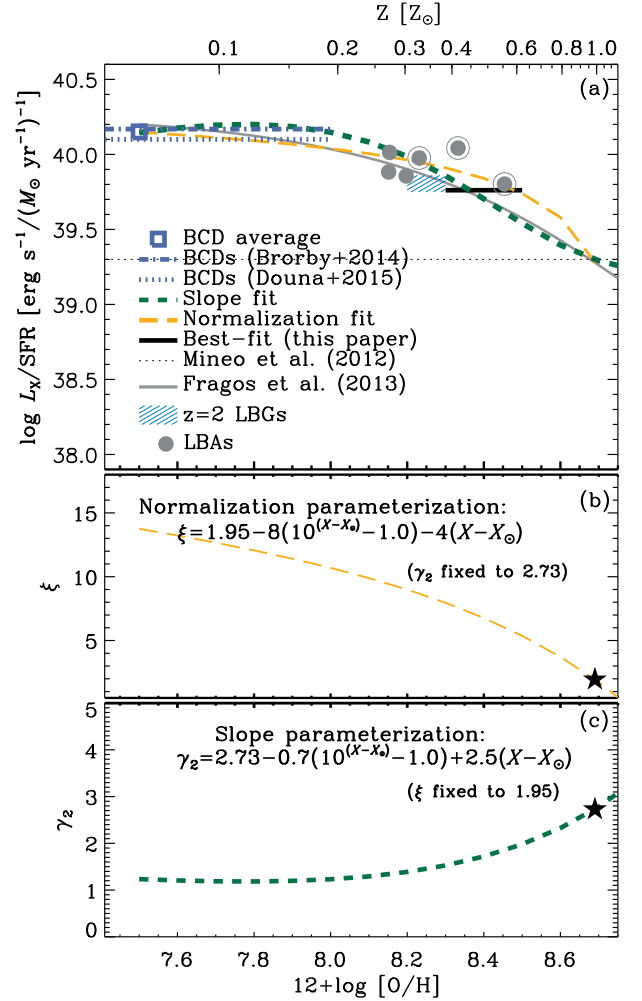}
    \end{center}
 \vspace{-0.15in}
  \caption{We parameterize the XLF normalization (green short-dashed) and slope (orange long-dashed) as functions of metallicity by fitting the observed \lhx/SFR of the full sample of 6 LBAs (gray points), the average BCD population (blue square) and the local star-forming galaxy relation (black dotted line) at solar metallicity (12+$\log$[O/H]=8.69). The top panel (a) is similar to Figures \ref{fig:metal} and \ref{fig:metalsim}, but has been expanded to include the extremely metal-poor dwarf galaxies, shown as blue lines (dash-dotted for \citealt{Brorby2014} and dotted for \citealt{Douna15}). We also included our best-fit XLF as a black solid line, spanning the metallicity range for Haro~11 and VV~114. The middle (b) and bottom (c) panels show the parameterizations of the normalization ($\xi$) and bright-end slope ($\gamma_2$), respectively. Each parameter is fit while holding the other fixed at the standard XLF value. The black stars show the standard XLF values at solar metallicity.}
   \label{fig:metalparam}
   \end{figure}

Next, we use the full sample of 6 LBAs from \citet{me-chandralba} to parameterize the values for $\gamma_2$ and the normalization, $\xi$ as functions of metallicity (12+$\log$[O/H]), matching to the standard XLF parameters at solar metallicity \citep[12+$\log$(O/H)=8.69;][]{solarmetal} and adding an average measurement from blue compact dwarfs (BCDs) at low metallicity based on studies by \citet{Brorby2014} and \citet{Douna15}. Explicitly, we integrate the XLFs given by the parameterized $\xi(X)$ and $\gamma_2(X)$, where $X\equiv12+\log$[O/H], down to $10^{36}$ \ergs to model \lhx/SFR for our parameterization and fit with the observed \lhx/SFR for all 6 LBAs. We fit these functional forms to the normalization (fixing the slope) and slope (fixing the normalization): 
\begin{equation}
\xi(X, \gamma_2=2.73)= 1.95+A(10^{X-X_\odot}-1.0) + B(X-X_\odot)\label{metalpar1}
\end{equation}
\begin{equation}
\gamma_2(X, \xi=1.95)= 2.73+C(10^{X-X_\odot}-1.0) + D(X-X_\odot)\label{metalpar2}
\end{equation}
where $X_\odot=8.69$. These functional forms follow the curve given by \citet{Fragos13}, which is a polynomial fit to metallicity in Z-units corresponding to our first term at first-order. However, we found that we also needed to add a linear term (in 12+$\log$[O/H] units) to follow the \lhx/SFR shape at higher metallicities in the simplest way. The best fits yield $A=-8^{+7}_{-8}$, $B=-4^{+7}_{-6}$, $C=-0.7^{+0.3}_{-0.4}$, $D=2.5\pm{0.4}$. The effects that these parameterizations have on \lhx/SFR are shown as orange long-dashed and green short-dashed lines in Figure \ref{fig:metalparam}. In the lower two panels, we show the functions of $\xi(X)$ [middle] and $\gamma_2(X)$ [lowest]. 
The black star marks the standard XLF at solar metallicity. We expanded the metallicity range to extend to the extremely low metallicity dwarf galaxies, studied by \citet[dash-dotted line]{Brorby2014} and \citet[dotted line]{Douna15}, shown as blue lines in the top panel, and the average value from these studies (shown by the blue square) was included in our fit. These studies used Bayesian probability methods to measure the normalization and slope of their sample's XLF, assuming the XLF is described by a single power law with a cut-off around $10^{40}$ \ergss. Although we did not include the LBGs (marked by the shaded cyan region) in our fits for the normalization and slope parameterization, our curves in the top panel show good agreement with their \lhx/SFR values.  

Our aims for parameterizing the normalization and slope of the XLF as a function of metallicity are two-fold: (1) to perform a consistency check with the integrated XLF properties in comparing observed values of \lhx/SFR for different galaxies as shown in the top panel of Figure \ref{fig:metalparam}, and (2) to provide the first-ever metallicity-dependent XLFs, given by Equations \ref{metalpar1} and \ref{metalpar2}. We remind readers that parameterizations mainly fit the $12+\log$[O/H]$=8.15$--8.69 range (including only the average of the low metallicity galaxies at $12+\log$[O/H]$=7.5$) and caution that this relation needs to be further tested for metallicities $12+\log$[O/H]$<7.5, between \approx7.5$--8.1, or $>$8.7. We find that the \citet{Fragos13} model can be approximated fairly well by a parameterization of the normalization and slope, even though the XRB population synthesis models do not themselves predict the shape of the XLF as a function of metallicity. 

\section{Conclusion and Future Work}\label{sec:conclusion}
We have studied the HMXB populations within two spatially-extended LBAs, analogs of high SFR galaxies from the early Universe. Previous studies have shown that LBAs exhibit elevated \lhx/SFR, which appears to be inversely correlated with metallicity according to X-ray binary population synthesis models \citep[\eg][]{Fragos13}. This paper focuses on the bright end of luminosity distribution of HMXBs in these low-metallicity star-forming galaxies to study how the HMXB populations within such galaxies compare with more typical (solar metallicity, on average) star-forming galaxies to address: can a different luminosity distribution (\eg higher normalization or flatter bright-end slope) of HMXBs within low-metallicity galaxies explain their elevated \lx/SFR compared to local star-forming galaxies? 

We find evidence that there is an increased number of high luminosity ($>10^{40}$\ergs) HMXBs. Since our galaxies have distances where source-blending plays a role, we account for these effects by simulating luminosity distributions using \HST~images as priors for the spatial distributions and the typical star-forming galaxy XLF \citep[referred to as the ``standard'' XLF; broken power law with normalization, $\xi=1.95$, $\gamma_1$=1.58 for \lhx$<1.1\times10^{40}$\ergs$\equiv L_b$, $\gamma_2=2.73$ above $L_b$;][]{M12} as the prior for the luminosity distribution. We tested different input XLFs by varying the normalizations and bright-end slopes ($\gamma_2$). We find that the standard XLF does not reproduce the observed data (ruled out at the 99\% confidence level by the KS test). Our best fit XLF (Model \# 7) has a flatter bright-end slope ($\gamma_2=1.90$) and higher normalization ($\xi=2.93$, 1.5$\times$ higher) than the standard. 

Based on optical and infrared emission line ratios, AGN do not contribute significantly to the global emission in Haro~11 \citep{Grimes07, Cormier2012} and VV~114 \citep{Grimes06}. However, we also removed all potential AGN sources from our analysis in a conservative attempt to test XLFs for that case. The best-fit simulated XLF to the data which excludes potential AGN (Model \#5) draws from an XLF with the same shape ($\gamma_2=2.73$) but higher normalization ($1.5\times$) than the standard. However, neither the standard XLF nor the previous best-fit XLF (Model \#7; with shallower bright-end slope, $\gamma_2=1.90$, and same normalization as Model \#5) can be ruled out at the 99\% confidence level by the KS test and Model \#7 appears to provide the next best-fit after Model \#5. However, without data at the bright-end (inhabited by the potential AGN) the models are not well-constrained.  

We parameterize the XLF normalizations and bright-end slopes with changing metallicity to fit \lhx/SFR for the full sample of LBAs, the average BCD, and the local star-forming population at solar metallicity. To avoid overfitting we only fit one parameter at a time, keeping the other fixed at the standard XLF value. For the normalization fit, we keep the bright-end slope fixed to $\gamma_2=2.73$, and for the bright-end slope fit we fix the normalization to $\xi=1.95$. We find that the fits (given by Equations 2 and 3) also describe the \lhx/SFR elevation observed in LBGs. 

Deeper observations of larger samples of low-metallicity, high SFR galaxies would offer better constraints on the XLFs that characterize HMXBs. In particular, deeper observations on low-metallicity galaxies provides more robust constraints on the XLF by probing fainter luminosities. Observing additional galaxies (more LBAs and metal-poor starbursts) can improve upon our results to differentiate better between models by sampling the high luminosity end of the XLF, where the difference between models is maximized.

Finally, future theoretical modeling of the luminosity distributions of HMXBs, for different metallicities and stellar population star-formation history scenarios, would provide meaningful comparisons with the observations. Recent advances in our understanding of uncertain phases of binary evolution such as the common envelope \citep[e.g.][]{Ivanova15,Nandez15}, and the availability of modern and computationally efficient detailed binary evolution codes \citep[e.g.][]{Paxton15} will soon enable population studies of ULXs that go beyond the approximations of parametric population synthesis codes (e.g. the implicit assumption of thermal stability of the stars and the simplistic common-envelope evolution prescriptions). These ``next generation'' population models will capture in much greater detail the physical effects of the varying metallicity and stellar age, in both the internal stellar structure and the evolution of the binary, and thus predict more accurately both the formation efficiency and the X-ray luminosity distribution of ULXs as a function of these two factors. Preliminary hybrid studies towards this direction, which combine approximate population synthesis calculations with grids of detailed binary calculations, already showed promising results for the study of a subset of ULXs, namely ULXs with neutron star accretors \citep{Fragos15b,Shao15}.

This research impacts our understanding of the high redshift Universe. The early Universe contained few metals, and galaxies forming at these epochs had high SFRs, thereby, producing numerous luminous HMXBs. It is likely the HMXBs produced in these high redshift galaxies contributed significantly to heating the Universe \citep{Mirabel11, Mesinger13, Pacucci14, Pober2015, Ryu15}. By understanding the HMXB populations in low-metallicity star-forming galaxies, locally, we can better interpret the X-ray emission from early generations of galaxies and estimate their significance. 

 \begin{acknowledgments}
We thank the referee for helpful suggestions that improved the manuscript.~A.R.B. and A.H. gratefully acknowledge the NASA Astrophysics Data Analysis Program (ADAP grant 09-ADP09-0071, PI: A. Hornschemeier) for providing financial support. T.F. acknowledges support from the Ambizione Fellowship of the Swiss National Science Foundation (grant PZ00P2\_148123).~A.Z. acknowledges funding from the European Research Council under the European UnionÕs Seventh Framework Programme (FP/2007-2013)/ERC Grant Agreement n. 617001 and financial support from NASA/ADAP grant NNX12AN05G.
\end{acknowledgments}

 \newcommand{\noop}[1]{}


\begin{thebibliography}{99}
\expandafter\ifx\csname natexlab\endcsname\relax\def\natexlab#1{#1}\fi

\bibitem[{{Adamo} {et~al.}(2010){Adamo}, {{\"O}stlin}, {Zackrisson}, {Hayes},
  {Cumming}, \& {Micheva}}]{Adamo2010}
{Adamo}, A., {et~al.} 2010, \mnras, 407, 870

\bibitem[{{Afonso} {et~al.}(2003){Afonso}, {Hopkins}, {Mobasher}, \&
  {Almeida}}]{Afonso2003}
{Afonso}, J., {Hopkins}, A., {Mobasher}, B., \& {Almeida}, C. 2003, \apj, 597,
  269

\bibitem[{{Asplund} {et~al.}(2009){Asplund}, {Grevesse}, {Sauval}, \&
  {Scott}}]{solarmetal}
{Asplund}, M., {Grevesse}, N., {Sauval}, A.~J., \& {Scott}, P. 2009, \araa, 47,
  481

\bibitem[{{Bachetti} {et~al.}(2013){Bachetti}, {Rana}, {Walton}, {Barret},
  {Harrison}, {Boggs}, {Christensen}, {Craig}, {Fabian}, {F{\"u}rst},
  {Grefenstette}, {Hailey}, {Hornschemeier}, {Madsen}, {Miller}, {Ptak},
  {Stern}, {Webb}, \& {Zhang}}]{Bachetti2013}
{Bachetti}, M., {et~al.} 2013, \apj, 778, 163

\bibitem[{{Baldwin} {et~al.}(1981){Baldwin}, {Phillips}, \& {Terlevich}}]{BPT}
{Baldwin}, J.~A., {Phillips}, M.~M., \& {Terlevich}, R. 1981, \pasp, 93, 5

\bibitem[{{Basu-Zych} {et~al.}(2013{\natexlab{a}}){Basu-Zych}, {Lehmer},
  {Hornschemeier}, {Bouwens}, {Fragos}, {Oesch}, {Belczynski}, {Brandt},
  {Kalogera}, {Luo}, {Miller}, {Mullaney}, {Tzanavaris}, {Xue}, \&
  {Zezas}}]{me-lbgstack}
{Basu-Zych}, A.~R., {et~al.} 2013{\natexlab{a}}, \apj, 762, 45

\bibitem[{{Basu-Zych} {et~al.}(2013{\natexlab{b}}){Basu-Zych}, {Lehmer},
  {Hornschemeier}, {Gon{\c c}alves}, {Fragos}, {Heckman}, {Overzier}, {Ptak},
  \& {Schiminovich}}]{me-chandralba}
---. 2013{\natexlab{b}}, \apj, 774, 152

\bibitem[{{Basu-Zych} {et~al.}(2009)}]{me-ifu}
{Basu-Zych}, A.~R. {et~al.} 2009, \apjl, 699, L118

\bibitem[{{Belczynski} {et~al.}(2008){Belczynski}, {Kalogera}, {Rasio}, {Taam},
  {Zezas}, {Bulik}, {Maccarone}, \& {Ivanova}}]{startrack}
{Belczynski}, K., {et~al.} 2008, \apjs, 174, 223

\bibitem[{{Bell}(2003)}]{Bell03}
{Bell}, E.~F. 2003, \apj, 586, 794

\bibitem[{{Bell} {et~al.}(2005){Bell}, {Papovich}, {Wolf}, {Le Floc'h},
  {Caldwell}, {Barden}, {Egami}, {McIntosh}, {Meisenheimer},
  {P{\'e}rez-Gonz{\'a}lez}, {Rieke}, {Rieke}, {Rigby}, \& {Rix}}]{Bell05}
{Bell}, E.~F., {et~al.} 2005, \apj, 625, 23

\bibitem[{{Binder} {et~al.}(2015){Binder}, {Gross}, {Williams}, \&
  {Simons}}]{Binder2015}
{Binder}, B., {Gross}, J., {Williams}, B.~F., \& {Simons}, D. 2015, \mnras,
  451, 4471

\bibitem[{{Bouwens} {et~al.}(2009)}]{Bouwens09}
{Bouwens}, R.~J. {et~al.} 2009, \apj, 705, 936

\bibitem[{{Boylan-Kolchin} {et~al.}(2009){Boylan-Kolchin}, {Springel}, {White},
  {Jenkins}, \& {Lemson}}]{MS-II}
{Boylan-Kolchin}, M., {Springel}, V., {White}, S.~D.~M., {Jenkins}, A., \&
  {Lemson}, G. 2009, \mnras, 398, 1150

\bibitem[{{Brandt} {et~al.}(2001)}]{BrandtLyBreak}
{Brandt}, W.~N. {et~al.} 2001, \apjl, 558, L5

\bibitem[{{Brorby} {et~al.}(2014){Brorby}, {Kaaret}, \&
  {Prestwich}}]{Brorby2014}
{Brorby}, M., {Kaaret}, P., \& {Prestwich}, A. 2014, \mnras, 441, 2346

\bibitem[{{Buat} {et~al.}(2005){Buat}, {Iglesias-P{\'a}ramo}, {Seibert},
  {Burgarella}, {Charlot}, {Martin}, {Xu}, {Heckman}, {Boissier}, {Boselli},
  {Barlow}, {Bianchi}, {Byun}, {Donas}, {Forster}, {Friedman}, {Jelinski},
  {Lee}, {Madore}, {Malina}, {Milliard}, {Morissey}, {Neff}, {Rich},
  {Schiminovitch}, {Siegmund}, {Small}, {Szalay}, {Welsh}, \&
  {Wyder}}]{Buat2005}
{Buat}, V., {et~al.} 2005, \apjl, 619, L51

\bibitem[{{Cash}(1979)}]{cstat}
{Cash}, W. 1979, \apj, 228, 939

\bibitem[{{Colbert} {et~al.}(2004){Colbert}, {Heckman}, {Ptak}, {Strickland},
  \& {Weaver}}]{Colbert04}
{Colbert}, E.~J.~M., {Heckman}, T.~M., {Ptak}, A.~F., {Strickland}, D.~K., \&
  {Weaver}, K.~A. 2004, \apj, 602, 231

\bibitem[{{Cormier} {et~al.}(2012){Cormier}, {Lebouteiller}, {Madden}, {Abel},
  {Hony}, {Galliano}, {Baes}, {Barlow}, {Cooray}, {De Looze}, {Galametz},
  {Karczewski}, {Parkin}, {R{\'e}my}, {Sauvage}, {Spinoglio}, {Wilson}, \&
  {Wu}}]{Cormier2012}
{Cormier}, D., {et~al.} 2012, \aap, 548, A20

\bibitem[{{Cowie} {et~al.}(2012){Cowie}, {Barger}, \& {Hasinger}}]{Cowie12}
{Cowie}, L.~L., {Barger}, A.~J., \& {Hasinger}, G. 2012, \apj, 748, 50

\bibitem[{{Crowther} {et~al.}(2010){Crowther}, {Barnard}, {Carpano}, {Clark},
  {Dhillon}, \& {Pollock}}]{Crowther2010}
{Crowther}, P.~A., {et~al.} 2010, \mnras, 403, L41

\bibitem[{{da Cunha} {et~al.}(2010){da Cunha}, {Eminian}, {Charlot}, \&
  {Blaizot}}]{daCunha2010}
{da Cunha}, E., {Eminian}, C., {Charlot}, S., \& {Blaizot}, J. 2010, \mnras,
  403, 1894

\bibitem[{{de Grijs} {et~al.}(2003{\natexlab{a}})}]{deGrijs03a}
{de Grijs}, R. {et~al.} 2003{\natexlab{a}}, \mnras, 342, 259

\bibitem[{{de Grijs} {et~al.}(2003{\natexlab{b}})}]{deGrijs03b}
---. 2003{\natexlab{b}}, \mnras, 343, 1285

\bibitem[{{Douna} {et~al.}(2015){Douna}, {Pellizza}, {Mirabel}, \&
  {Pedrosa}}]{Douna15}
{Douna}, V.~M., {Pellizza}, L.~J., {Mirabel}, I.~F., \& {Pedrosa}, S.~E. 2015,
  ArXiv e-prints

\bibitem[{{Erb} {et~al.}(2006)}]{erb06}
{Erb}, D.~K. {et~al.} 2006, \apj, 644, 813

\bibitem[{{Fragos} {et~al.}(2012){Fragos}, {Lehmer}, {Tremmel}, {Tzanavaris},
  {Basu-Zych}, {Belczynski}, {Hornschemeier}, {Jenkins}, {Kalogera}, {Ptak}, \&
  {Zezas}}]{F12}
{Fragos}, T., {et~al.} 2012, ArXiv e-prints

\bibitem[{{Fragos} {et~al.}(\noop{2013a}2013a){Fragos}, {Lehmer}, {Tremmel},
  {Tzanavaris}, {Basu-Zych}, {Belczynski}, {Hornschemeier}, {Jenkins},
  {Kalogera}, {Ptak}, \& {Zezas}}]{Fragos12}
---. \noop{2013a}2013a, \apj, 764, 41

\bibitem[{{Fragos} {et~al.}(\noop{2013b}2013b){Fragos}, {Lehmer}, {Naoz},
  {Zezas}, \& {Basu-Zych}}]{Fragos13}
{Fragos}, T., {Lehmer}, B.~D., {Naoz}, S., {Zezas}, A., \& {Basu-Zych}, A.
  \noop{2013b}2013b, \apjl, 776, L31

\bibitem[{{Fragos} {et~al.}(2015){Fragos}, {Linden}, {Kalogera}, \&
  {Sklias}}]{Fragos15b}
{Fragos}, T., {Linden}, T., {Kalogera}, V., \& {Sklias}, P. 2015, \apjl, 802,
  L5

\bibitem[{{Fragos} \& {McClintock}(2015)}]{Fragos15a}
{Fragos}, T. \& {McClintock}, J.~E. 2015, \apj, 800, 17

\bibitem[{{Garn} \& {Best}(2010)}]{Garn2010}
{Garn}, T. \& {Best}, P.~N. 2010, \mnras, 409, 421

\bibitem[{{Giavalisco}(2002)}]{giarev}
{Giavalisco}, M. 2002, \araa, 40, 579

\bibitem[{{Gladstone} {et~al.}(2009){Gladstone}, {Roberts}, \&
  {Done}}]{Gladstone09}
{Gladstone}, J.~C., {Roberts}, T.~P., \& {Done}, C. 2009, \mnras, 397, 1836

\bibitem[{{Grimes} {et~al.}(2006)}]{Grimes06}
{Grimes}, J.~P. {et~al.} 2006, \apj, 648, 310

\bibitem[{{Grimes} {et~al.}(2007)}]{Grimes07}
---. 2007, \apj, 668, 891

\bibitem[{{Heckman} {et~al.}(2005)}]{Heckman05}
{Heckman}, T.~M. {et~al.} 2005, \apjl, 619, L35

\bibitem[{{Hoopes} {et~al.}(2007)}]{choopes}
{Hoopes}, C.~G. {et~al.} 2007, \apjs, 173, 441

\bibitem[{{Hopkins} {et~al.}(2001){Hopkins}, {Connolly}, {Haarsma}, \&
  {Cram}}]{Hopkins2001}
{Hopkins}, A.~M., {Connolly}, A.~J., {Haarsma}, D.~B., \& {Cram}, L.~E. 2001,
  \aj, 122, 288

\bibitem[{{Iglesias-P{\'a}ramo} {et~al.}(2007){Iglesias-P{\'a}ramo}, {Buat},
  {Hern{\'a}ndez-Fern{\'a}ndez}, {Xu}, {Burgarella}, {Takeuchi}, {Boselli},
  {Shupe}, {Rowan-Robinson}, {Babbedge}, {Conrow}, {Fang}, {Farrah},
  {Gonz{\'a}lez-Solares}, {Lonsdale}, {Smith}, {Surace}, {Barlow}, {Forster},
  {Friedman}, {Martin}, {Morrissey}, {Neff}, {Schiminovich}, {Seibert},
  {Small}, {Wyder}, {Bianchi}, {Donas}, {Heckman}, {Lee}, {Madore}, {Milliard},
  {Rich}, {Szalay}, {Welsh}, \& {Yi}}]{IglesiasParamo07}
{Iglesias-P{\'a}ramo}, J., {et~al.} 2007, \apj, 670, 279

\bibitem[{{Iono} {et~al.}(2013){Iono}, {Saito}, {Yun}, {Kawabe}, {Espada},
  {Hagiwara}, {Imanishi}, {Izumi}, {Kohno}, {Motohara}, {Nakanishi}, {Sugai},
  {Tateuchi}, {Tamura}, {Ueda}, \& {Yoshii}}]{Iono13}
{Iono}, D., {et~al.} 2013, \pasj, 65, L7

\bibitem[{{Ivanova} {et~al.}(2015){Ivanova}, {Justham}, \&
  {Podsiadlowski}}]{Ivanova15}
{Ivanova}, N., {Justham}, S., \& {Podsiadlowski}, P. 2015, \mnras, 447, 2181

\bibitem[{{James} {et~al.}(2013){James}, {Tsamis}, {Walsh}, {Barlow}, \&
  {Westmoquette}}]{James2013}
{James}, B.~L., {Tsamis}, Y.~G., {Walsh}, J.~R., {Barlow}, M.~J., \&
  {Westmoquette}, M.~S. 2013, \mnras, 430, 2097

\bibitem[{{Jia} {et~al.}(2011){Jia}, {Ptak}, {Heckman}, {Overzier},
  {Hornschemeier}, \& {LaMassa}}]{Jia}
{Jia}, J., {et~al.} 2011, \apj, 731, 55

\bibitem[{{Justham} {et~al.}(2015){Justham}, {Peng}, \&
  {Schawinski}}]{Justham15}
{Justham}, S., {Peng}, E.~W., \& {Schawinski}, K. 2015, \apjl, 809, L16

\bibitem[{{Kaaret}(2014)}]{kaaret2014}
{Kaaret}, P. 2014, ArXiv e-prints

\bibitem[{{Kaaret} {et~al.}(2011{\natexlab{a}}){Kaaret}, {Schmitt}, \&
  {Gorski}}]{kaaret}
{Kaaret}, P., {Schmitt}, J., \& {Gorski}, M. 2011{\natexlab{a}}, \apj, 741, 10

\bibitem[{{Kaaret} {et~al.}(2011{\natexlab{b}}){Kaaret}, {Schmitt}, \&
  {Gorski}}]{Kaaret2011}
---. 2011{\natexlab{b}}, \apj, 741, 10

\bibitem[{{Kauffmann} {et~al.}(2003){Kauffmann}, {Heckman}, {Tremonti},
  {Brinchmann}, {Charlot}, {White}, {Ridgway}, {Brinkmann}, {Fukugita}, {Hall},
  {Ivezi{\'c}}, {Richards}, \& {Schneider}}]{Kauffmann03AGN}
{Kauffmann}, G., {et~al.} 2003, \mnras, 346, 1055

\bibitem[{{Kewley} \& {Ellison}(2008)}]{KewleyEllison08}
{Kewley}, L.~J. \& {Ellison}, S.~L. 2008, \apj, 681, 1183

\bibitem[{{Kroupa}(2001)}]{Kroupa}
{Kroupa}, P. 2001, \mnras, 322, 231

\bibitem[{{Laird} {et~al.}(2006){Laird}, {Nandra}, {Hobbs}, \&
  {Steidel}}]{Laird06}
{Laird}, E.~S., {Nandra}, K., {Hobbs}, A., \& {Steidel}, C.~C. 2006, \mnras,
  373, 217

\bibitem[{{Laycock} {et~al.}(2015{\natexlab{a}}){Laycock}, {Cappallo}, \&
  {Moro}}]{Laycock2015a}
{Laycock}, S.~G.~T., {Cappallo}, R.~C., \& {Moro}, M.~J. 2015{\natexlab{a}},
  \mnras, 446, 1399

\bibitem[{{Laycock} {et~al.}(2015{\natexlab{b}}){Laycock}, {Maccarone}, \&
  {Christodoulou}}]{Laycock2015b}
{Laycock}, S.~G.~T., {Maccarone}, T.~J., \& {Christodoulou}, D.~M.
  2015{\natexlab{b}}, \mnras, 452, L31

\bibitem[{{Lehmer} {et~al.}(2005){Lehmer}, {Brandt}, {Alexander}, {Bauer},
  {Conselice}, {Dickinson}, {Giavalisco}, {Grogin}, {Koekemoer}, {Lee},
  {Moustakas}, \& {Schneider}}]{Lehmer05}
{Lehmer}, B.~D., {et~al.} 2005, \aj, 129, 1

\bibitem[{{Lehmer} {et~al.}(2015){Lehmer}, {Tyler}, {Hornschemeier}, {Wik},
  {Yukita}, {Antoniou}, {Boggs}, {Christensen}, {Craig}, {Hailey}, {Harrison},
  {Maccarone}, {Ptak}, {Stern}, {Zezas}, \& {Zhang}}]{Lehmer15}
---. 2015, \apj, 806, 126

\bibitem[{{Lehmer} {et~al.}(2013){Lehmer}, {Wik}, {Hornschemeier}, {Ptak},
  {Antoniou}, {Argo}, {Bechtol}, {Boggs}, {Christensen}, {Craig}, {Hailey},
  {Harrison}, {Krivonos}, {Leyder}, {Maccarone}, {Stern}, {Venters}, {Zezas},
  \& {Zhang}}]{Lehmer13}
---. 2013, \apj, 771, 134

\bibitem[{{Lehmer} {et~al.}(2008)}]{Lehmer08}
{Lehmer}, B.~D. {et~al.} 2008, \apj, 681, 1163

\bibitem[{{Lehmer} {et~al.}(2010)}]{Lehmer2010}
---. 2010, \apj, 724, 559

\bibitem[{{Linden} {et~al.}(2010){Linden}, {Kalogera}, {Sepinsky}, {Prestwich},
  {Zezas}, \& {Gallagher}}]{Linden2010}
{Linden}, T., {et~al.} 2010, \apj, 725, 1984

\bibitem[{{Mapelli} {et~al.}(2009){Mapelli}, {Colpi}, \&
  {Zampieri}}]{Mapelli09}
{Mapelli}, M., {Colpi}, M., \& {Zampieri}, L. 2009, \mnras, 395, L71

\bibitem[{{Mapelli} {et~al.}(2011){Mapelli}, {Ripamonti}, {Zampieri}, \&
  {Colpi}}]{Mapelli2011}
{Mapelli}, M., {Ripamonti}, E., {Zampieri}, L., \& {Colpi}, M. 2011,
  Astronomische Nachrichten, 332, 414

\bibitem[{{Mapelli} {et~al.}(2010){Mapelli}, {Ripamonti}, {Zampieri}, {Colpi},
  \& {Bressan}}]{Mapelli2010}
{Mapelli}, M., {Ripamonti}, E., {Zampieri}, L., {Colpi}, M., \& {Bressan}, A.
  2010, \mnras, 408, 234

\bibitem[{{Mesinger} {et~al.}(2013){Mesinger}, {Ferrara}, \&
  {Spiegel}}]{Mesinger13}
{Mesinger}, A., {Ferrara}, A., \& {Spiegel}, D.~S. 2013, \mnras, 431, 621

\bibitem[{{Mineo} {et~al.}(2012){Mineo}, {Gilfanov}, \& {Sunyaev}}]{M12}
{Mineo}, S., {Gilfanov}, M., \& {Sunyaev}, R. 2012, \mnras, 419, 2095

\bibitem[{{Mineo} {et~al.}(2014){Mineo}, {Rappaport}, {Levine}, {Pooley},
  {Steinhorn}, \& {Homan}}]{Mineo14}
{Mineo}, S., {et~al.} 2014, \apj, 797, 91

\bibitem[{{Mirabel} {et~al.}(2011){Mirabel}, {Dijkstra}, {Laurent}, {Loeb}, \&
  {Pritchard}}]{Mirabel11}
{Mirabel}, I.~F., {Dijkstra}, M., {Laurent}, P., {Loeb}, A., \& {Pritchard},
  J.~R. 2011, \aap, 528, A149

\bibitem[{{Nandez} {et~al.}(2015){Nandez}, {Ivanova}, \& {Lombardi}}]{Nandez15}
{Nandez}, J.~L.~A., {Ivanova}, N., \& {Lombardi}, J.~C. 2015, \mnras, 450, L39

\bibitem[{{Nandra} {et~al.}(2002)}]{Nandra02}
{Nandra}, K. {et~al.} 2002, \apj, 576, 625

\bibitem[{{Overzier} {et~al.}(2008)}]{rod}
{Overzier}, R.~A. {et~al.} 2008, \apj, 677, 37

\bibitem[{{Pacucci} {et~al.}(2014){Pacucci}, {Mesinger}, {Mineo}, \&
  {Ferrara}}]{Pacucci14}
{Pacucci}, F., {Mesinger}, A., {Mineo}, S., \& {Ferrara}, A. 2014, \mnras, 443,
  678

\bibitem[{{Paxton} {et~al.}(2015){Paxton}, {Marchant}, {Schwab}, {Bauer},
  {Bildsten}, {Cantiello}, {Dessart}, {Farmer}, {Hu}, {Langer}, {Townsend},
  {Townsley}, \& {Timmes}}]{Paxton15}
{Paxton}, B., {et~al.} 2015, ArXiv e-prints

\bibitem[{{Pettini} \& {Pagel}(2004)}]{PP04}
{Pettini}, M. \& {Pagel}, B.~E.~J. 2004, \mnras, 348, L59

\bibitem[{{Pober} {et~al.}(2015){Pober}, {Ali}, {Parsons}, {McQuinn},
  {Aguirre}, {Bernardi}, {Bradley}, {Carilli}, {Cheng}, {DeBoer}, {Dexter},
  {Furlanetto}, {Grobbelaar}, {Horrell}, {Jacobs}, {Klima}, {Kohn}, {Liu},
  {MacMahon}, {Maree}, {Mesinger}, {Moore}, {Razavi-Ghods}, {Stefan},
  {Walbrugh}, {Walker}, \& {Zheng}}]{Pober2015}
{Pober}, J.~C., {et~al.} 2015, ArXiv e-prints

\bibitem[{{Podsiadlowski} {et~al.}(2003){Podsiadlowski}, {Rappaport}, \&
  {Han}}]{Pod03}
{Podsiadlowski}, P., {Rappaport}, S., \& {Han}, Z. 2003, \mnras, 341, 385

\bibitem[{{Prestwich} {et~al.}(2015){Prestwich}, {Jackson}, {Kaaret}, {Brorby},
  {Roberts}, {Saar}, \& {Yukita}}]{PrestwichHaro11}
{Prestwich}, A.~H., {et~al.} 2015, ArXiv e-prints

\bibitem[{{Prestwich} {et~al.}(2007){Prestwich}, {Kilgard}, {Crowther},
  {Carpano}, {Pollock}, {Zezas}, {Saar}, {Roberts}, \& {Ward}}]{Prestwich07}
---. 2007, \apjl, 669, L21

\bibitem[{{Prestwich} {et~al.}(2013){Prestwich}, {Tsantaki}, {Zezas},
  {Jackson}, {Roberts}, {Foltz}, {Linden}, \& {Kalogera}}]{Prestwich2013}
---. 2013, \apj, 769, 92

\bibitem[{{Rana} {et~al.}(2015){Rana}, {Harrison}, {Bachetti}, {Walton},
  {Furst}, {Barret}, {Miller}, {Fabian}, {Boggs}, {Christensen}, {Craig},
  {Grefenstette}, {Hailey}, {Madsen}, {Ptak}, {Stern}, {Webb}, \&
  {Zhang}}]{Rana15}
{Rana}, V., {et~al.} 2015, \apj, 799, 121

\bibitem[{{Rappaport} {et~al.}(2005){Rappaport}, {Podsiadlowski}, \&
  {Pfahl}}]{Rapp05}
{Rappaport}, S.~A., {Podsiadlowski}, P., \& {Pfahl}, E. 2005, \mnras, 356, 401

\bibitem[{{Reddy} {et~al.}(2008){Reddy}, {Steidel}, {Pettini}, {Adelberger},
  {Shapley}, {Erb}, \& {Dickinson}}]{Reddy2008}
{Reddy}, N.~A., {et~al.} 2008, \apjs, 175, 48

\bibitem[{{Ryu} {et~al.}(2015){Ryu}, {Tanaka}, \& {Perna}}]{Ryu15}
{Ryu}, T., {Tanaka}, T.~L., \& {Perna}, R. 2015, ArXiv e-prints

\bibitem[{{Saito} {et~al.}(2015){Saito}, {Iono}, {Yun}, {Ueda}, {Nakanishi},
  {Sugai}, {Espada}, {Imanishi}, {Motohara}, {Hagiwara}, {Tateuchi}, {Lee}, \&
  {Kawabe}}]{Saito2015}
{Saito}, T., {et~al.} 2015, \apj, 803, 60

\bibitem[{Schmitt {et~al.}(2006)Schmitt, Calzetti, Armus, Giavalisco, Heckman,
  Kennicutt, Leitherer, \& Meurer}]{Schmitt2006}
Schmitt, H.~R., {et~al.} 2006, \apjs, 164, 52

\bibitem[{{Seibert} {et~al.}(2002){Seibert}, {Heckman}, \&
  {Meurer}}]{Seibert2002}
{Seibert}, M., {Heckman}, T.~M., \& {Meurer}, G.~R. 2002, \aj, 124, 46

\bibitem[{{Shao} \& {Li}(2015)}]{Shao15}
{Shao}, Y. \& {Li}, X.-D. 2015, \apj, 802, 131

\bibitem[{{Taam} \& {Sandquist}(2000)}]{TS2000}
{Taam}, R.~E. \& {Sandquist}, E.~L. 2000, \araa, 38, 113

\bibitem[{{Tremonti} {et~al.}(2004)}]{Tremonti}
{Tremonti}, C.~A. {et~al.} 2004, \apj, 613, 898

\bibitem[{{Valsecchi} {et~al.}(2010){Valsecchi}, {Glebbeek}, {Farr}, {Fragos},
  {Willems}, {Orosz}, {Liu}, \& {Kalogera}}]{Val2010}
{Valsecchi}, F., {et~al.} 2010, \nat, 468, 77

\bibitem[{{Veilleux} \& {Osterbrock}(1987)}]{VO87}
{Veilleux}, S. \& {Osterbrock}, D.~E. 1987, \apjs, 63, 295

\bibitem[{{Walton} {et~al.}(2013){Walton}, {Fuerst}, {Harrison}, {Stern},
  {Bachetti}, {Barret}, {Bauer}, {Boggs}, {Christensen}, {Craig}, {Fabian},
  {Grefenstette}, {Hailey}, {Madsen}, {Miller}, {Ptak}, {Rana}, {Webb}, \&
  {Zhang}}]{Walton13}
{Walton}, D.~J., {et~al.} 2013, \apj, 779, 148

\bibitem[{{Walton} {et~al.}(2014){Walton}, {Harrison}, {Grefenstette},
  {Miller}, {Bachetti}, {Barret}, {Boggs}, {Christensen}, {Craig}, {Fabian},
  {Fuerst}, {Hailey}, {Madsen}, {Parker}, {Ptak}, {Rana}, {Stern}, {Webb}, \&
  {Zhang}}]{Walton14}
---. 2014, \apj, 793, 21

\bibitem[{{Wang} \& {Heckman}(1996)}]{Wang96}
{Wang}, B. \& {Heckman}, T.~M. 1996, \apj, 457, 645

\bibitem[{{Wong} {et~al.}(2014){Wong}, {Valsecchi}, {Ansari}, {Fragos},
  {Glebbeek}, {Kalogera}, \& {McClintock}}]{Wong2014}
{Wong}, T.-W., {et~al.} 2014, \apj, 790, 119

\bibitem[{{Wong} {et~al.}(2012){Wong}, {Valsecchi}, {Fragos}, \&
  {Kalogera}}]{Wong2012}
{Wong}, T.-W., {Valsecchi}, F., {Fragos}, T., \& {Kalogera}, V. 2012, \apj,
  747, 111

\bibitem[{{Xue} {et~al.}(2011){Xue}, {Luo}, {Brandt}, {Bauer}, {Lehmer},
  {Broos}, {Schneider}, {Alexander}, {Brusa}, {Comastri}, {Fabian}, {Gilli},
  {Hasinger}, {Hornschemeier}, {Koekemoer}, {Liu}, {Mainieri}, {Paolillo},
  {Rafferty}, {Rosati}, {Shemmer}, {Silverman}, {Smail}, {Tozzi}, \&
  {Vignali}}]{Xue11}
{Xue}, Y.~Q., {et~al.} 2011, \apjs, 195, 10

\bibitem[{{Zezas} {et~al.}(2006){Zezas}, {Fabbiano}, {Baldi}, {Schweizer},
  {King}, {Ponman}, \& {Rots}}]{Zezas06}
{Zezas}, A., {et~al.} 2006, \apjs, 166, 211

\bibitem[{{Zezas} {et~al.}(2007){Zezas}, {Fabbiano}, {Baldi}, {Schweizer},
  {King}, {Rots}, \& {Ponman}}]{Zezas_antennae}
---. 2007, \apj, 661, 135

\end{thebibliography}
\end{document}